\documentclass[journal]{IEEEtran}
\newcommand{\VMAFp}{$\text{VMAF}\mathrm{_p}$}
\newcommand{\SSIMp}{$\text{SSIM}\mathrm{_p}$}
\newcommand{\MSIMp}{$\text{MS-SSIM}\mathrm{_p}$}

\usepackage{footnote}
\makesavenoteenv{tabular}
\usepackage{amsfonts} %For \mathbb
\usepackage[dvipsnames]{xcolor}
\usepackage[normalem]{ulem}
\usepackage{cite}
\ifCLASSINFOpdf
  \usepackage[pdftex]{graphicx}
\else
\fi
\usepackage{amsmath}
\newcommand\norm[1]{\left\lVert#1\right\rVert}
\newcommand\ul[1]{\underline{#1}}
\usepackage{array}
\usepackage{tabularx,booktabs}
\usepackage{multirow}
\ifCLASSOPTIONcompsoc
  \usepackage[caption=false,font=normalsize,labelfont=sf,textfont=sf]{subfig}
\else
  \usepackage[caption=false,font=footnotesize]{subfig}
\fi
\begin{document}
\title{ProxIQA: A Proxy Approach to Perceptual Optimization of Learned Image Compression}

\author{Li-Heng~Chen,
        Christos~G.~Bampis,
        Zhi~Li,
        Andrey~Norkin,
        and~Alan~C.~Bovik,~\IEEEmembership{Fellow,~IEEE}% <-this % stops a space
        
\thanks{L.~Chen and A.~C.~Bovik are with the Department of Electrical and Computer Engineering, University of Texas at Austin, Austin, TX, 78712 USA (email:lhchen@utexas.edu, bovik@ece.utexas.edu).}% <-this % stops a space
\thanks{C.~G.~Bampis, Z.~Li, and A. Norkin are with Netflix Inc. Los Gatos, CA, 95032 USA (email:christosb@netflix.com, zli@netflix.com, anorkin@netflix.com).}% <-this % stops a space
\thanks{This work is supported by Netflix Inc.}}

% \markboth{Journal of \LaTeX\ Class Files,~Vol.~14, No.~8, August~2015}%
% {Shell \MakeLowercase{\textit{et al.}}: Bare Demo of IEEEtran.cls for IEEE Journals}



\maketitle
\begin{abstract}
The use of $\ell_p$ (p=1,2) norms has largely dominated the measurement of loss  in neural networks due to their simplicity and analytical properties. However, when used to assess the loss of visual information, these simple norms are not very consistent with human perception. Here, we describe a different ``proximal" approach to optimize image analysis networks against quantitative perceptual models. Specifically, we construct a proxy network, broadly termed ProxIQA, which mimics the perceptual model while serving as a loss layer of the network. We experimentally demonstrate how this optimization framework can be applied to train an end-to-end optimized image compression network. By building on top of an existing deep image compression model, we are able to demonstrate a bitrate reduction of as much as 31\% over MSE optimization, given a specified perceptual quality (VMAF) level.
\end{abstract}


% \fi
\IEEEpeerreviewmaketitle

\section{Introduction}
\IEEEPARstart{R}{ecently} deep neural networks have been successfully and ubiquitously applied on diverse image processing and computer vision tasks, such as semantic segmentation \cite{ShelhamerLD17}, object recognition \cite{HeZRS16}, video encoding \cite{TMLiu2018, SPaul2020}, and optical flow \cite{Dosovitskiy:2015:FLO}. Many classic image transformation problems can be approached using a deep generative network, which learns to reconstruct high-quality output images from degraded input image(s). Explicitly, the generative network is trained in a supervised manner with a loss function, which is used to measure the fidelity between the output and a ground-truth image. For instance, the denoising task aims to reconstruct a noise-free image from a noisy image, and Convolutional Neural Networks (CNNs) have been shown to provide good noisy-to-pristine mapping functions \cite{BurgerSH12,NIPS2008_3506}. Similar tasks where retaining image fidelity is important include deep image compression, super-resolution \cite{dong2014learning,MSLapSRN}, frame interpolation \cite{liu2019cyclicgen}, and so on.

Although a significant amount of research has been applied on deep learning image transformation problems, most of this work has focused on investigating network architectures or improving convergence speed. The selection of an appropriate loss function, however, has not been studied as much. The choice of the loss functions used to guide model training has been largely limited to the $\ell_p$ norm family, in particular the MSE (squared $\ell_2$ norm), the $\ell_1$ norm, and variants of these\cite{BarronCVPR2019}. The structural similarity quality index (SSIM) \cite{WangBSS04} and its multi-scale version (MS-SSIM) \cite {WangMSSSIM03} have also been adopted as loss functions for several image reconstruction tasks \cite{Snell2017,Zhao2017}, owing to their perceptual relevance and good analytic properties, such as differentiability.

\begin{figure}[!t]
    \centering
    \includegraphics[width=3.49in]{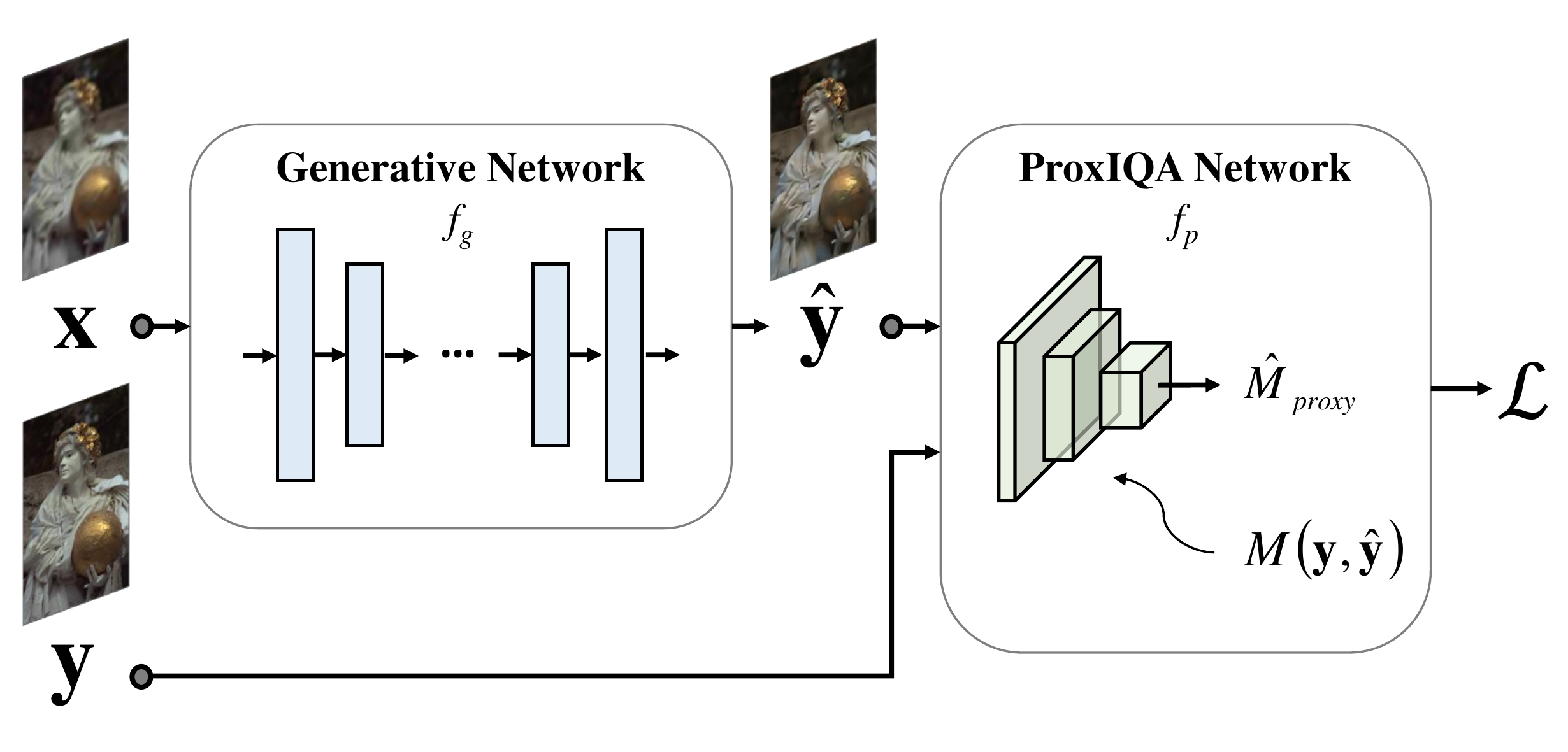}
    \caption{General framework of perceptual optimization using a ProxIQA network: A generative network takes an image $\mathbf{x}$ as input and outputs a reconstructed image $\hat{\mathbf{y}}=f_g(\mathbf{x};\theta_g)$. Note that $\theta_g, \theta_p$ are the parameters of $f_g,f_p$, respectively, while $\mathbf{y}$ is the ground-truth image. Given an image quality measurement $M$, the ProxIQA network is learned as its proxy, where the output $\hat{M}_\mathrm{proxy}$ represents $M(\mathbf{y},\hat{\mathbf{y}})$.}
    \label{fig_prox_framework}
\end{figure}

Perceptual image quality assessment has been a long-standing research problem. Although numerous powerful perceptual models have been proposed to predict the  perceived quality of a distorted picture, most other image quality indexes have never been adopted as deep network loss functions, because they are generally \textit{non-differentiable} and functionally complex.

Towards bridging the gap between modern perceptual quality models and deep generative networks, we explore the potential of adapting more powerful and sophisticated perceptual image quality models as loss functions in deep neural network for addressing the aforementioned problems, by simulating the measurements made by a perceptual model by a proxy network ProxIQA. As shown in Fig. \ref{fig_prox_framework}, the main idea is to optimize the hyper parameters of the generative network $\theta_g$, using a ProxIQA network as a perceptual loss function
\begin{equation}
\begin{split}
  \mathcal{L}\left( \theta_g \right) & =f_p\left( \mathbf{y},\hat{\mathbf{y}};\theta_p \right)\\
             & =f_p\left( \mathbf{y},f_g\left( \mathbf{x};\theta_g\right);\theta_p \right),
\end{split}
\end{equation}
where $\mathbf{x}$, $\hat{\mathbf{y}}$ are the input and output of the generative network $f_g$ and $\mathbf{y}$ is the ground-truth image. In the image compression problem, $\mathbf{x}$ is an uncompressed image and we anticipate the fidelity of the compressed image. Thus, this is a special case where $\mathbf{x}=\mathbf{y}$. The parameters $\theta_p$ of the ProxIQA network are optimized so that it mimics $M(\mathbf{y},\hat{\mathbf{y}})$, a perceptual image quality measurement between $\mathbf{y}$ and $\hat{\mathbf{y}}$.
 
The outline of this paper is as follows: Section II reviews the relevant literature of image quality assessment, perceptual optimization, and deep image compression. Section III describes the ProxIQA framework, while Section IV provides analysis and experimental results. Finally, Section V concludes the paper.

%% You must have at least 2 lines in the paragraph with the drop %letter
%% (should never be an issue)
%I wish you the best of success.
\begin{figure}[!t]
\centering
\subfloat[Determined function loss $\mathcal{L}_{func}$]{\includegraphics[width=2.5in]{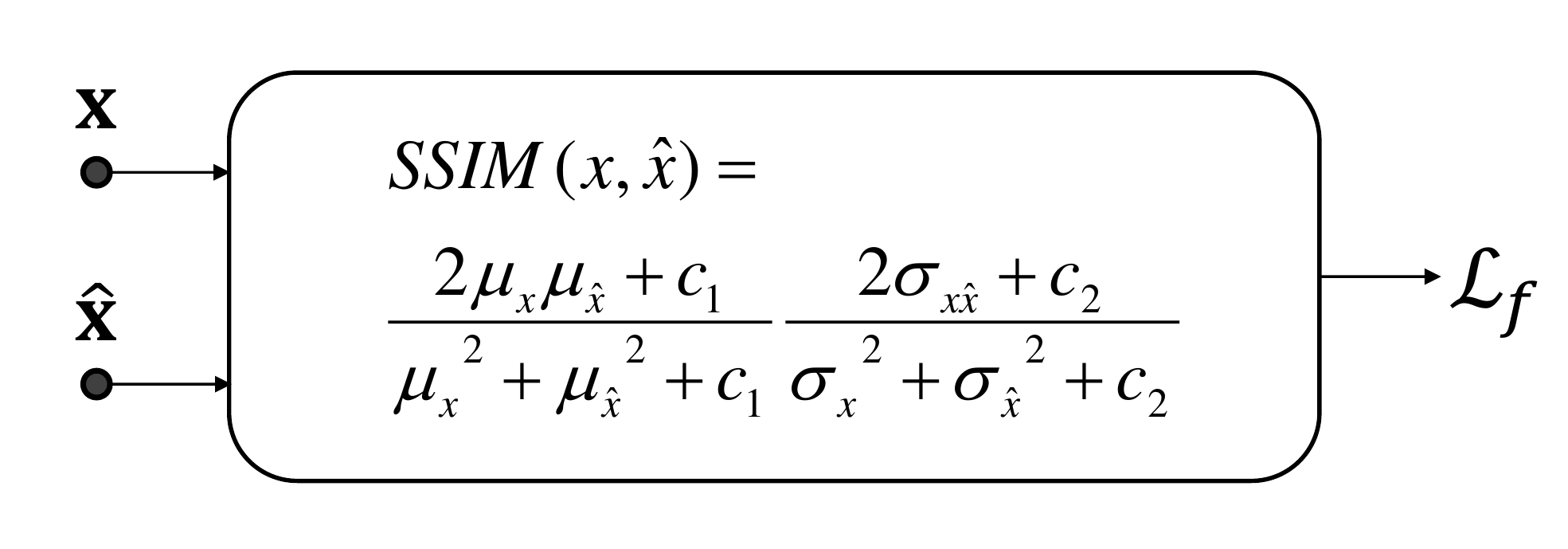}%
\label{fig_comp_a}}
\hfil
\subfloat[Perceptual loss $\mathcal{L}_{perc}$]{\includegraphics[width=2.5in]{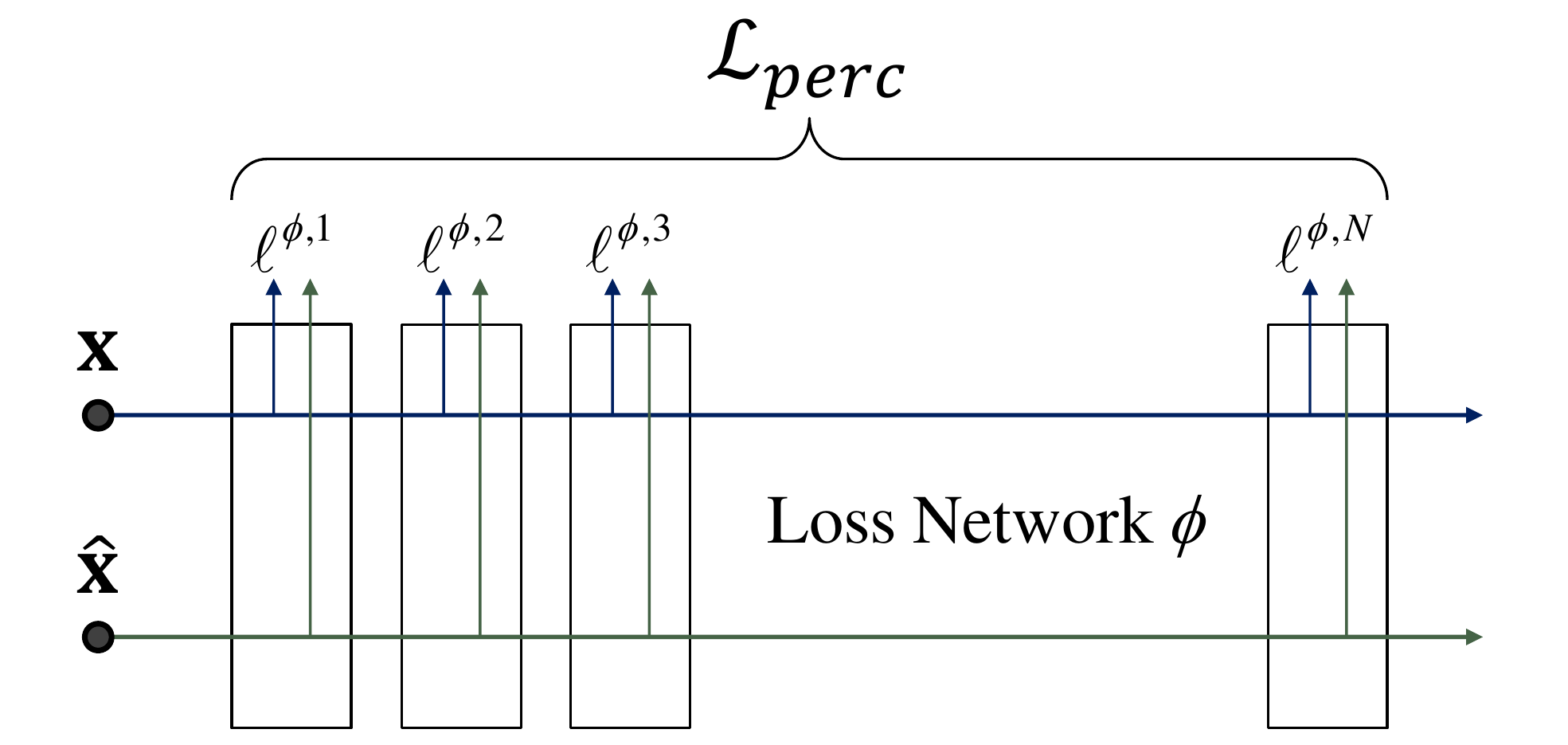}%
\label{fig_comp_b}}
\hfil
\subfloat[ProxIQA loss $\mathcal{L}_{prox}$]{\includegraphics[width=2.5in]{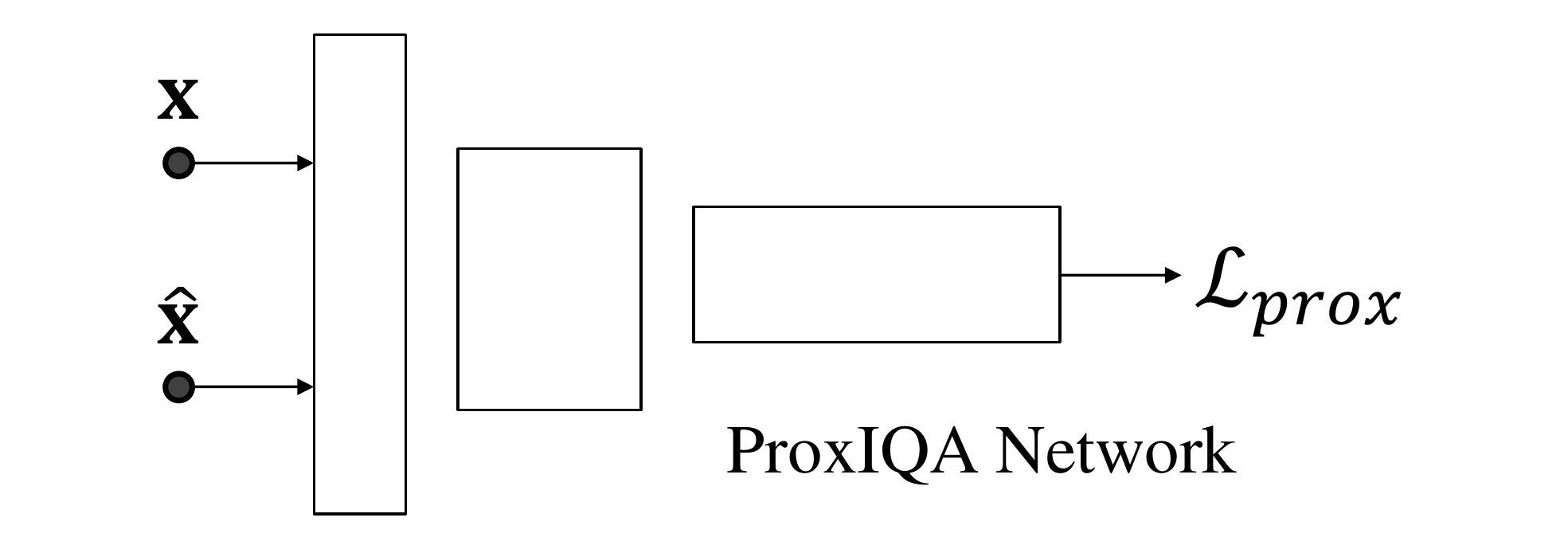}%
\label{fig_comp_c}}
\caption{Comparison of different perceptual loss layers for generative neural networks. Given a target patch $\mathbf{x}$ and a reconstructed patch $\hat{\mathbf{x}}$, (a) determined function based approaches typically use a differentiable function having a certain degree of convexity, such as SSIM and MS-SSIM. (b) Perceptual-loss based approaches define the loss from the features extracted from intermediate layers of a trained network (such as VGG) (c) Our method uses the output of a proxy network that approximates an IQA model as the loss function.}
\label{fig_sim}
\end{figure}

\section{Related Work}
In this section, we provide a literature review of studies that are closely related to our work. The relevant topics of objective image quality assessment, perceptual optimization and deep compression are briefly reviewed. 
\subsection{Perceptual Image Quality Metrics}
Over the past decade there has been a remarkably increasing interest in developing objective image quality assessment (IQA) methods. Objective IQA models are commonly classified as full-reference, reduced-reference \cite{Wan2020}, and no-reference \cite{ztuugcvqa2020}, based on the amount of information they assess from a reference image of ostensibly pristine quality. Here we only need to consider the full-reference (FR) scenario, since it may be assumed that ground-truth data is available, hence we only review FR IQA models.

Beyond the well-known structural similarity index and other SSIM-type methods, a wide variety of perception-based FR models have been designed, including the visual signal-to-noise ratio index (VSNR) \cite{DMCSSHVSNR07}, the visual information fidelity (VIF) index \cite{SheikhB06}, the MAD model \cite{Chandler2010}, the feature similarity index (FSIM) and its extension FSIM$_c$ \cite{LZhangFSIM2011}, and the Visual Saliency-Induced index (VSI) \cite{ZhangSLVSI14}.

With the rapid development of machine learning, important data-driven models have also begun to emerge. Models using ``handcrafted" features with different regressors \cite{Liu2013,PeiDOG15,LukinPIEA15,Oszust2016} can produce accurate quality predictions on existing datasets. However, they cannot be used for end-to-end training if any component is not differentiable, motivating us to pursue a systematic and holistic approach. For example, Random Forest regression as used by \cite{PeiDOG15} is non-differentiable at the split nodes. Other DNN-based IQA models \cite{GaoWLTYZ17,Bosse2018, Bosse2019dsp} have also shown promising results. A particularly successful example is Netflix's announcement of an open-source FR video quality engine called Video Multimethod Assessment Fusion (VMAF) \cite{ZliVMAF18}. VMAF combines multiple quality features to train a Support Vector Regressor (SVR) to predict subjective judgments of video quality. When it is applied to still pictures, VMAF treats the data as a video frame having zero motion. The study in \cite{Sinno2020} showed that VMAF correlates quite well against human judgements of still picture quality. Indeed, most of the VMAF features are those of VIF \cite{SheikhB06}, which is among the most powerful and effective image quality models. Like SSIM, VMAF is used to perceptually optimize tremendous volumes of internet picture and video traffic \cite{ZliVMAF18}.

Generally, more advanced, high-performance quality prediction models such as these are difficult to adopt as loss functions for end-to-end optimization networks.

\begin{figure*}[!t]
    \centering
    \includegraphics[width=6in]{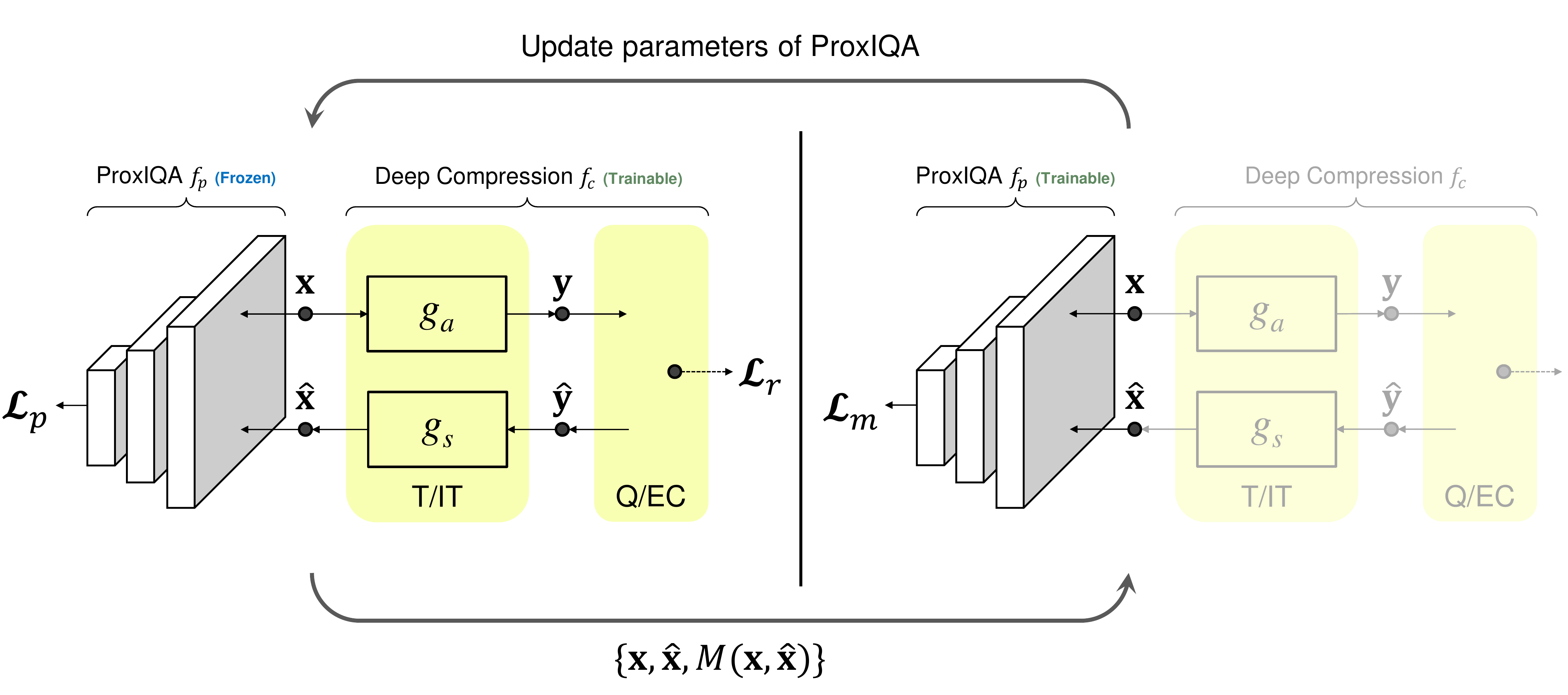}
    \caption{Detailed framework of the proposed optimization strategy. Perceptually training a deep image compression model involves alternating optimization of the compression network (left side of the figure) and the ProxIQA network (right side of the figure). Thin arrows indicate the flow of data in the network, while bold arrows represent the information being delivered to update the complementary network.}
    \label{fig_prox_bls}
\end{figure*}

\subsection{Perceptual Optimization}
As tractable tools for perceptual optimization, SSIM and MS-SSIM have been widely adopted because of the simple analytical form of their gradients and computational ease. Moreover, their convexity properties \cite{Brunet2012} makes them feasible targets for optimization. For example, two recent studies adopted structural similarity functions as loss layers of image generation models, obtaining improved results as validated by conducting a human subjective study \cite{Snell2017} and by objective evaluation against several other perceptual models \cite{Zhao2017}. The recent manuscript \cite{dingperceptualiqa2020} systematically evaluated IQA models as loss functions for perceptually optimizing different tasks. It was found that many IQA methods failed due to non-convexity of the objective function.

Rather than optimizing a mathematical function, another approach uses a deep neural network to guide the training. Recent experimental studies suggest that the features extracted from a well-trained image classification network have the capability to capture information useful for other perceptual tasks \cite{zhang2018perceptual}. As illustrated in Fig.~\ref{fig_sim}\subref{fig_comp_b}, the perceptual loss is defined as
\begin{equation}
\begin{split}
  \mathcal{L}_{perc} & =\sum_i\ell^{\phi,i} \\
             & =\sum_iN_i^{-1} \norm{ \phi_i\left(\mathbf{x}\right)-\phi_i\left(\mathbf{\hat{x}}\right) }_2^2,
\end{split}
\end{equation}
where $\phi_i$ denotes the output feature map of the $i$-th layer with $N_i$ elements of a pre-trained network $\phi$.

In practice, the loss computed from the high-level features extracted from a pre-trained VGG classification network \cite{SimonyanZ14a}, also called VGG loss, has been commonly adopted for diverse computer vision tasks. The VGG loss has been applied to such diverse tasks as style transfer \cite{JohnsonAF16,GatysEBHS17}, superresolution \cite{BrunaSL15,JohnsonAF16,LedigTHCCAATTWS17,Sajjadi2017}, and image inpainting \cite{Yang_2017_CVPR}. 

\subsection{End-to-end Optimized Lossy Image Compression}
Recently, lossy image compression models have been realized using deep neural network architectures. Most of these have deployed deep auto-encoders. For example, Ball\'e \textit{et al.} \cite{BalleLS16a} proposed a general infrastructure for optimizing image compression in an end-to-end manner. Unlike other methods, the bitrate is estimated and considered during training. In \cite{balle2018variational}, this model is improved by incorporating a scale hyperprior into the compression framework. The authors use an additional network to estimate the standard deviation of the quantized coefficients to further improve coding efficiency. Later, Minnen \textit{et al.} \cite{NIPS2018_8275} exploit a PixelCNN layer, which they combine with an autoregressive hyperprior. Beyond these early efforts, other recent approaches have adopted more complex network architectures such as recurrent neural networks (RNNs) \cite{Toderici2015VariableRI,Toderici2017,Johnston_2018_CVPR} and generative adversarial networks (GANs) \cite{agustsson2018generative,Lhdefink2019GANVJ}. Some works has also been done to extend these ideas to the deep video compression problem \cite{wu2018vcii,cheng19, kimvideo}.

Unsurprisingly, the idea of optimizing a conventional codec such as H.264/AVC against perceptual models like SSIM, VIF, and VMAF have been deeply studied \cite{channappayya08,YHH10,WangRWMG12, kslussim20} and implemented in widespread practice \cite{ZliVMAF18}. We seek to extend this concept in similar manner to learn an end-to-end perceptually optimized compression model.

%\IEEEpubidadjcol

\section{Proposed Perceptual Optimization Framework}
Our approach to training an image compression model in a perceptually optimized way is depicted in Fig.~\ref{fig_prox_bls}. This framework involves optimizing two networks: an image compression network $f_c$, and a sub-network $f_p$, which is a proxy of an IQA model, which we will refer to as ProxIQA. A source image $x$ is input to a compression network, which produces a reconstructed image:
\begin{equation}
  \hat{x} =f_c\left( x \right).
\end{equation}
Separately, the ProxIQA network maps the image pair $\left(x,\hat{x}\right)$ into a proxy of an image quality model $M$:
\begin{equation}
  \hat{M} =f_p\left( x,\hat{x} \right).
\end{equation}
In each training iteration, the two networks are alternately updated as follows:

\subsubsection{Deep Compression Model Updating} To integrate $f_p$ into the update of $f_c$ given a mini-batch $\mathbf{x}$, the model parameters of $f_p$ are fixed during training. In order to minimize distortion, the output of $f_p$ becomes part of the objective in the optimization of $f_c$:
\begin{equation}
  \hat{M}\left( \mathbf{x},\hat{\mathbf{x}} \right)=f_p\left( \mathbf{x},\hat{\mathbf{x}} \right) = f_p\left( \mathbf{x},f_c\left( \mathbf{x} \right)\right).
\end{equation}
By back-propagating through the forward model, the loss derivative is used to drive $f_c$. As we will see later in section \ref{sec:secIII}, the entropy model is also updated at this stage.

\subsubsection{ProxIQA Network Updating} Given a mini-batch pair $\mathbf{x}$ and $\hat{\mathbf{x}}$ collected from the most recent update of the compression network, the quality scores ${M}(\mathbf{x}, \hat{\mathbf{x}})$ are calculated. The ProxIQA network is updated such that its output $\hat{M}$ optimally fit ${M}$ given the input $\left\{\mathbf{x},\hat{\mathbf{x}}\right\}$. Note that the compression network is not needed in this part of the training.

As may be seen, the auxiliary sub-network ProxIQA is incorporated into the training of the compression network. However, it is important to understand that the ProxIQA network is not present during the testing (image compression/decompression) phase. In addition, it could be argued that the training strategy employed by ProxIQA shares some similarities with that of Generative Adversarial Networks (GANs) \cite{NIPSGAN}. However, we have identified an important conceptual difference between the two: the ProxIQA network learns to fit a quality model and does not include the image compression network during its update, which is different from the discriminator in GANs. 

\begin{figure}[!t]
  \centering
  \includegraphics[width=3.3in]{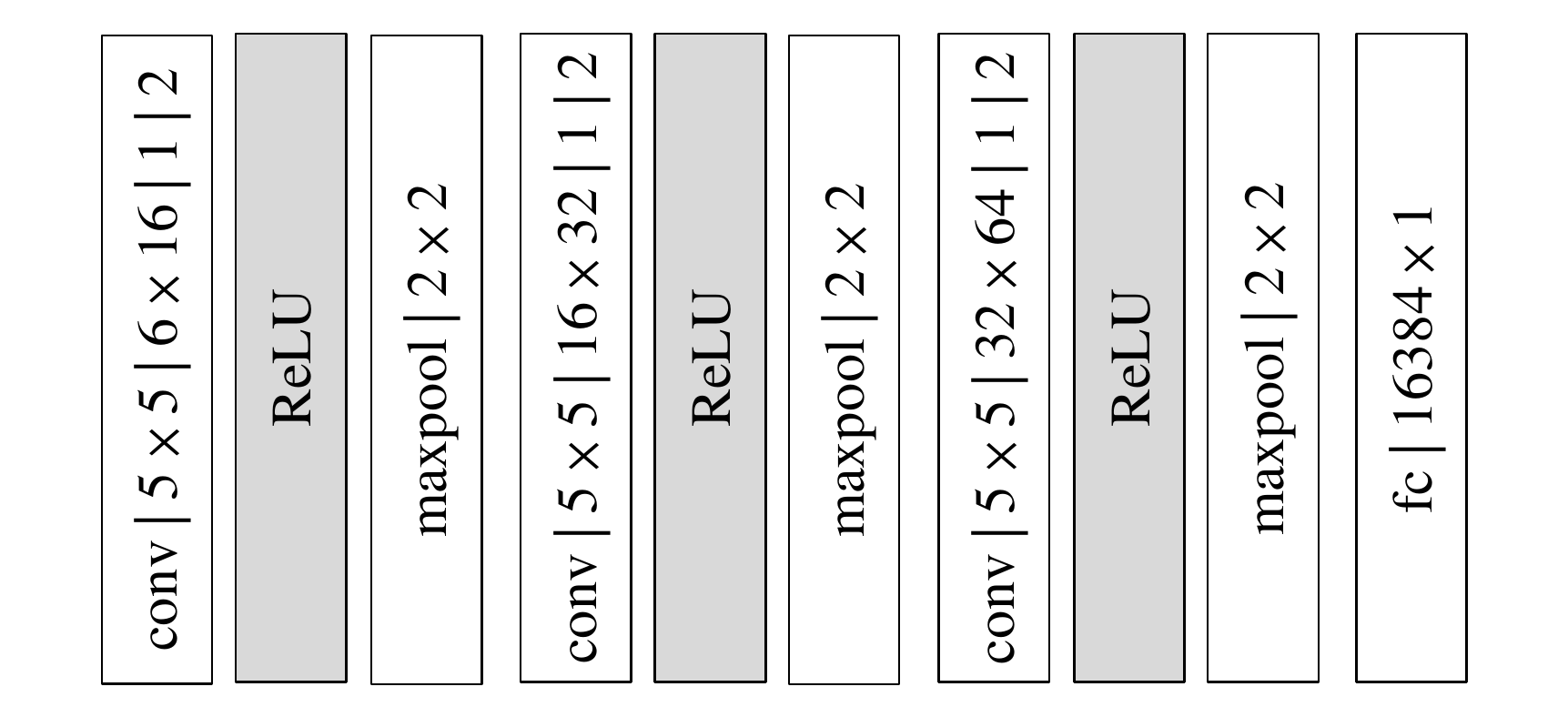}
  \caption{Architecture of the ProxIQA network. The convolutional parameters are denoted as: height $\times$ width $\mid$ input channel $\times$ output channel $\mid$ stride $\mid$ padding; The max pooling layers are denoted as: vertical pooling size $\times$ horizontal pooling size.}
  \label{fig_network}
\end{figure}

\subsection{Network Architecture}
\subsubsection{ProxIQA Network} The goal is to learn a nonlinear regressor via a CNN. The network takes a reference patch $x$ and a distorted patch $\hat{x}$ as input, where both have $W\times H$ pixels. They are then concatenated into a 6-channel signal, where a $W \times H \times 6$ raw input $\{x,\hat{x}\}$ is fed into the network and reduced to a predicted quality score. As depicted in Fig.~\ref{fig_network}, the ProxIQA network may be as simple as a shallow CNN consisting of three stages of convolution, ReLU as an activation function, and maxpooling. The spatial size is reduced by a factor of $2$ after each stage via $2\times2$ max pooling layers. Finally, $64$ $W/8\times H/8$ feature maps are flattened and fed to a fully connected layer which yields the output. The parameterization of each layer is detailed in the figure. 

\subsubsection{Compression Network} We build on the deep image compression model \cite{BalleLS16a}. As shown in Fig.~\ref{fig_prox_bls}, the image compression network comprises an analysis transform ($g_a$) at the encoder side, and a synthesis transform ($g_s$) at the decoder side. Both transforms are composed of three consecutive layers of convolution-down(up) sampling-activation. Instead of utilizing ReLU, a generalized divisive normalization (GDN) transform is adopted as the activation function \cite{pcs_Balle18}, similar to normalization of the visual signal by the human visual system (HVS).

\subsection{Loss Functions}\label{sec:loss}
\label{sec:secIII}
As illustrated in Fig.~\ref{fig_prox_bls}, let $\mathbf{x}$, $\mathbf{y}$, $\hat{\mathbf{y}}$, and $\hat{\mathbf{x}}$ be the source batch, latent presentation, quantized latent presentation, and reconstructed batch, respectively. The model parameters in the analysis and synthesis transforms are collectively denoted by $\theta=(\theta_a,\theta_s)$. The ProxIQA network has model parameters $\phi$. Given a perceptual metric $M$, such as SSIM or VMAF, the goal is to optimize the full set of parameters $\theta$, $\phi$, such that the learned image codec can generate a reconstructed image $\hat{\mathbf{x}}$ that has a high perceptual quality score $M(\mathbf{x}, \hat{\mathbf{x}})$. Furthermore, the rate should be as small as possible. Therefore, we train our model using the following losses.

\subsubsection{Rate loss} The rate loss representing the bit consumption of an encode $\hat{\mathbf{y}}$ is defined by
\begin{equation}\label{eq:rate_loss}
  \mathcal{L}_{r}\left(\theta \right)
  = -\log_2p_{\hat{\mathbf{y}}}\left( \hat{\mathbf{y}} \right),
\end{equation}
where $p_{\hat{\mathbf{y}}}\left( \hat{\mathbf{y}} \right)$ is the entropy model. This entropy term is minimized when the actual marginal distribution and $\hat{\mathbf{y}}$ have the same distribution.

During training, the latent presentation $\mathbf{y}$ is approximately quantized to $\hat{\mathbf{y}}$ as additive quantization noise $\Delta\mathbf{y}$ (i.i.d uniform). Then, $\hat{\mathbf{y}}$ is used to estimate the rate via (\ref{eq:rate_loss}). Unlike the estimated entropy used when training the network, a variant of the context-adaptive binary arithmetic coder (CABAC) \cite{cabac_tcsvt03} is used to encode the discrete-valued data into the bitstream during testing.

\subsubsection{Pixel loss} The pixel loss is the residual between the source image  and the reconstructed image mapped by a distance function $d$. Given $\mathbf{x}$ and $\hat{\mathbf{x}}$, the pixel loss is defined by
\begin{equation}\label{eq:pixel_loss}
  \mathcal{L}_{d}\left(\theta\right)
  =d\left( \mathbf{x}-\hat{\mathbf{x}} \right),
\end{equation}
where $d(.)$ can be mean squared error (i.e., $d(x)=\norm{x}_2^2$) or mean absolute error (i.e., $d(x)=\norm{x}_1$).

The original work in \cite{BalleLS16a} used $d(x)=\norm{x}_2^2$ as the pixel loss to maximize the PSNR of the reconstructed images. When combined with the rate loss, the image compression network is optimized by minimizing the objective function defined by
\begin{equation}\label{eq:baseline_loss}
  \mathcal{L}\left(\theta\right)=\lambda \mathcal{L}_{d} +\mathcal{L}_{r},
\end{equation}
which is a similar form of rate-distortion optimization (RDO) function as is found in conventional codecs. This equation indicates that the entropy model is jointly optimized with the compression network. We make use of a pixel loss to encourage training stability. 

\begin{figure*}[!t]
\centering
\subfloat[Reconstructed patches from deep compression model.]{\includegraphics[width=2in]{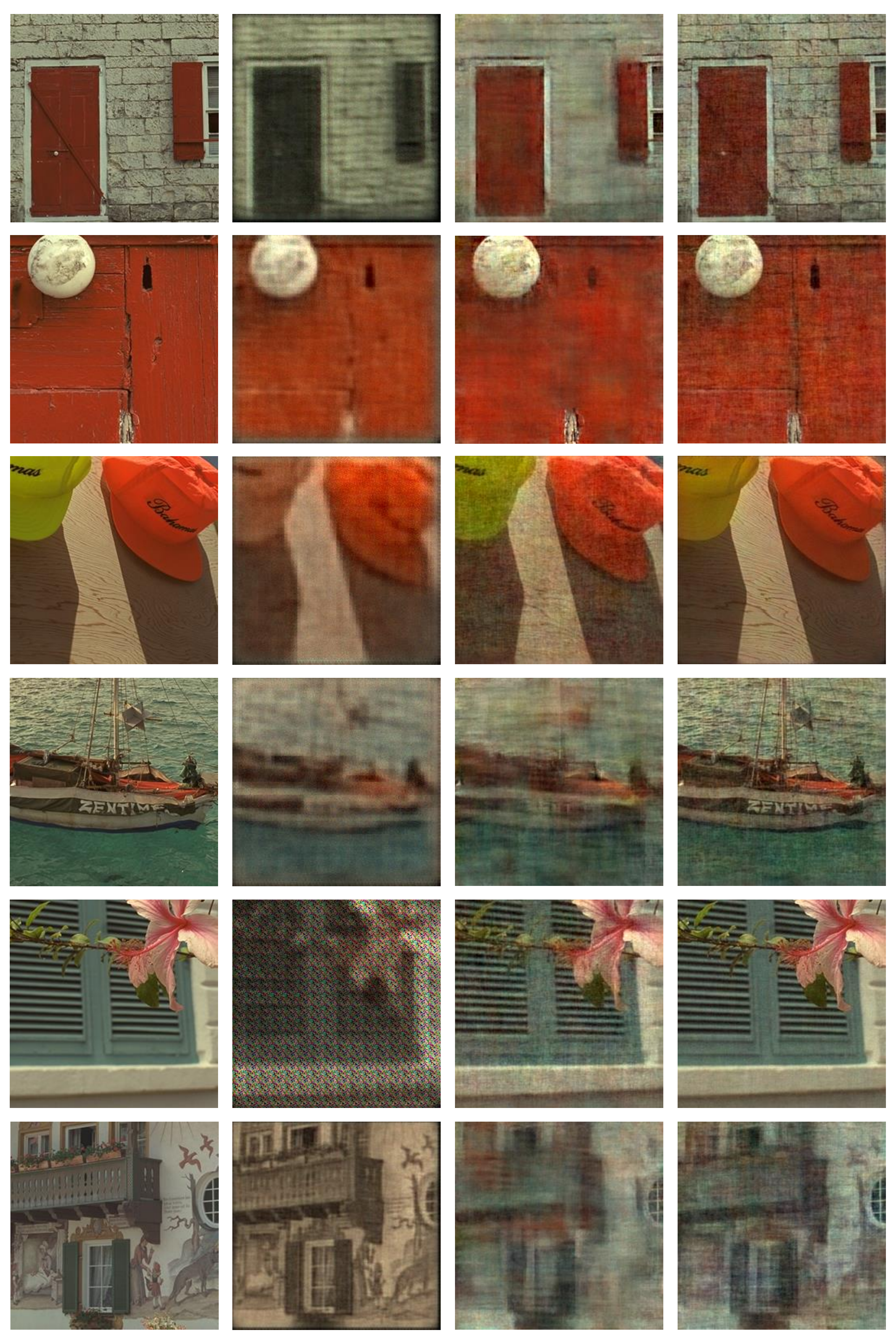}%
\label{fig_dataset_a}}
\hfil
\subfloat[Waterloo Exploration database.]
{\includegraphics[width=2in]{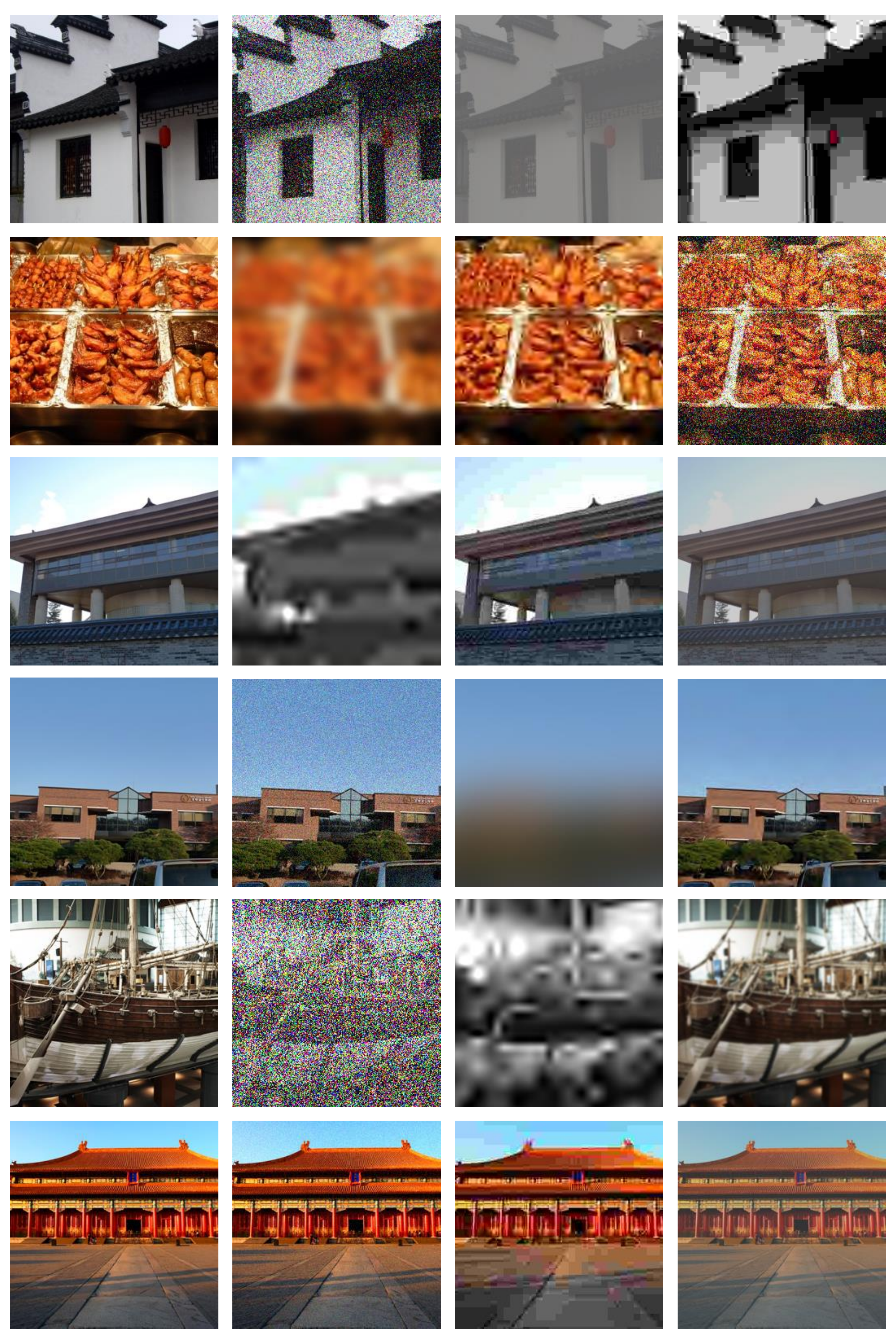}%
\label{fig_dataset_b}}
\hfil
\subfloat[BAPPS database.]
{\includegraphics[width=2in]{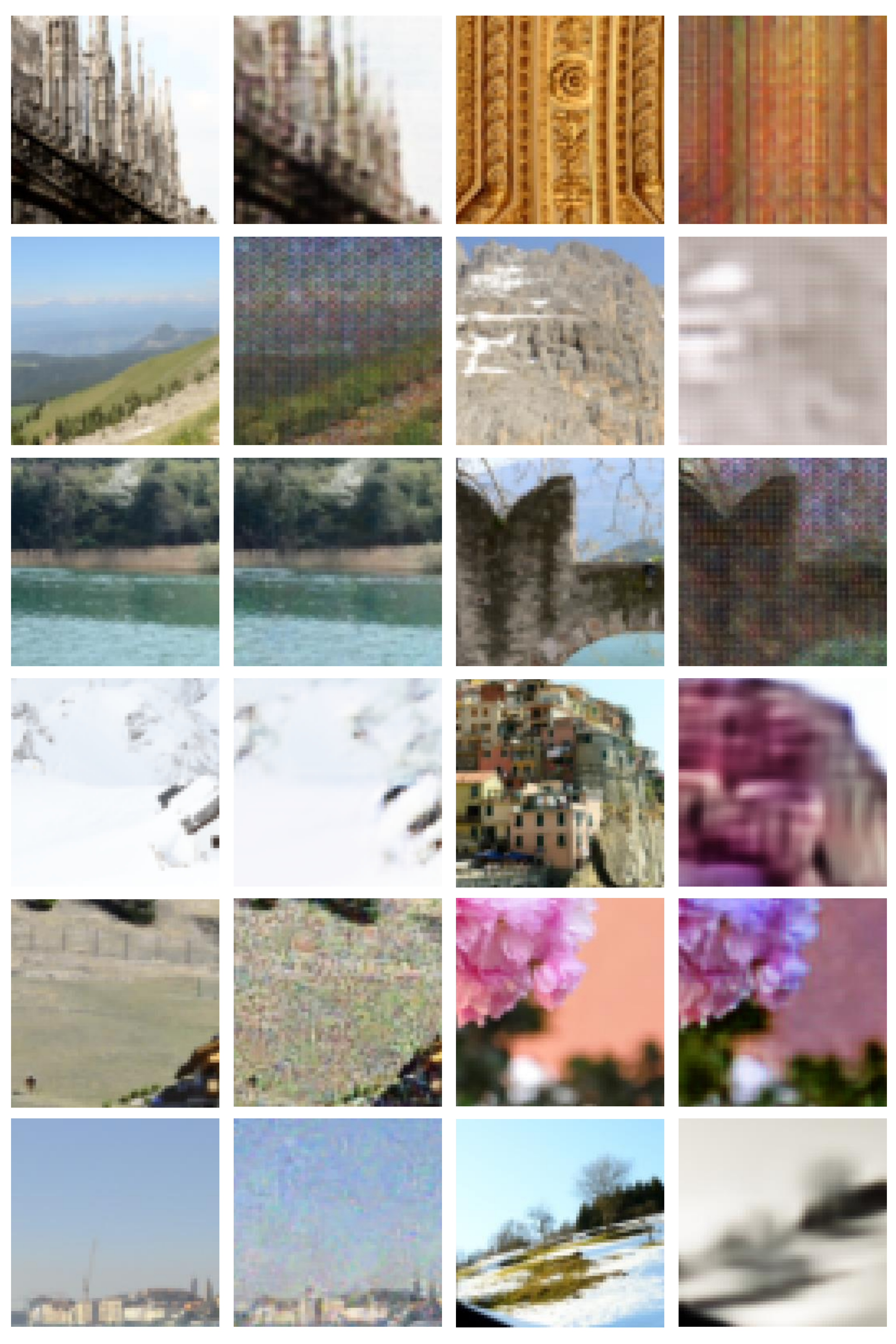}%
\label{fig_dataset_c}}
\caption{Comparison of distorted patches from existing databases and three deep compression networks. For (a)(b), The first column shows the pristine patches while the other three columns are randomly selected distorted patches. In (c), each similar pair includes a reference patch (left), and a distorted patch (right). The distortions in (a)(b)(c) were generated from (i) patches reconstructed during different training steps of the deep compression network; (ii) synthetic distortions with different severity levels added to the original patches; (iii) the outputs of various convolutional neural networks, respectively}.
\label{fig_dataset}
\end{figure*}

\subsubsection{Proxy loss} Instead of just minimizing an $\ell_p$ norm between $\mathbf{x}$ and $\hat{\mathbf{x}}$, we introduce a novel loss term. The proxy loss $\mathcal{L}_{p}$ is calculated from the output of ProxIQA network, denoted by $\hat{M}$, with fixed parameter $\phi$:
\begin{equation}\label{eq:proxy_loss}
  \mathcal{L}_{p}\left(\theta ; \phi\right)
  =M_\mathrm{max} - \hat{M}\left( \mathbf{x},\hat{\mathbf{x}} \right).
\end{equation}
Here $M_\mathrm{max}$ denotes the upper bound of the model $M$, which is a constant to the loss function.

Finally, the total loss function for optimizing the compression network is the weighted combination of the losses from Eqs. (\ref{eq:pixel_loss}), (\ref{eq:rate_loss}), and (\ref{eq:proxy_loss}):
\begin{equation}\label{eq:total_loss}
  \mathcal{L}_{t}\left(\theta ; \phi\right)
  =\lambda\left[ \alpha\mathcal{L}_{p} + \left( 1-\alpha \right) \mathcal{L}_{d} \right] + \mathcal{L}_{r},
\end{equation}
where $\lambda$ balances bitrate against distortion of the encoded bitstream, and $\alpha$ weights the proxy loss against the pixel loss. We empirically set $\alpha=1.5e-3$ for $M=\text{VMAF}$, and $\alpha=3e-3$ for both $M=\text{SSIM}$ and MS-SSIM. This was accomplished by selecting four values spaced along the reals and choosing the best performing value. However, the performance did not vary much over the range of tested values. The pixel loss plays a different role as a regularization term. Since the ProxIQA network is updated at each step, the loss function $\mathcal{L}_{p}$ is also updated. The pixel loss serves to stabilize the training process. 

\subsubsection{Metric loss} The ProxIQA network aims to mimic an image quality model $M$. Given two images $x$ and $\hat{x}$, define a metric loss $\mathcal{L}_{m}$ to attain this objective while updating ProxIQA network:
\begin{equation}\label{eq:metric_loss}
  \mathcal{L}_{m}\left(\phi;\hat{\mathbf{x}}\right)
  = \norm{ \hat{M}\left( \mathbf{x},\hat{\mathbf{x}} \right) - 
  M\left( \mathbf{x},\hat{\mathbf{x}} \right) }_2^2.
\end{equation}
Note that $\hat{\mathbf{x}}$ is a constant, since it is obtained from the reconstructed patches generated during the most recent update of the compression network. $M\left( \mathbf{x},\hat{\mathbf{x}} \right)$ is the ground-truth data obtained on-the-fly during training. We used the MSE loss, because it is more sensitive to outliers. We also tried $\ell_1$ loss to update the ProxIQA parameters $\phi$, obtaining very similar results.

\subsection{What's Wrong with Using a Pre-trained Network?}
Another way of attempting to accomplish the same goal is to use a pre-trained network as the loss layer. That is, a proxy network is first learned to predict a metric score given a pristine patch and a distorted patch from an existing dataset. Next, the trained proxy network is inserted into the loss layer of the deep compression network with the goal of maximizing the proxy score. Unfortunately, severe complications can arise when applying this simple methodology.

\subsubsection{New Distortion Types}
The success of a CNN model depends highly on the size of the training set. This is often an obstacle to learning DNN-based IQA models, due to the insufficient size, as compared to image recognition databases, of publicly available IQA databases. Luckily, training a proxy network on an existing model does not require human-labeled subjective quality scores such as mean opinion scores (MOS) or differential mean opinion scores (DMOS). The ground truth metric score for training the proxy network is easily obtained, given a pristine patch and a distorted patch. Therefore, we can make use of large-scale databases that do not include MOS, such as the Waterloo Exploration database \cite{ma2017waterloo}.

Nevertheless, the distortion types provided by public-domain databases are generally quite different from the distortions created by a deep compression model (see Fig.~\ref{fig_dataset}\subref{fig_dataset_a}). In fact, most existing databases only provide synthetic distortions. As shown in Fig.~\ref{fig_dataset}\subref{fig_dataset_b}, we randomly selected images distorted by JPEG, JPEG2000, Gaussian blur, and white Gaussian noise, from the Waterloo Exploration database. As may be easily observed, these distortions are drastically different from those created by CNNs.

The Berkeley-Adobe PPS (BAPPS) database \cite{zhang2018perceptual} contains many distorted patches collected from the outputs of CNN models. However, these CNN-based distortion types are still different from the patches reconstructed by deep compression networks. In the second to fourth columns of Fig.~\ref{fig_dataset}\subref{fig_dataset_a}, we show several randomly collected reconstructed patches output by the deep compression model during different training steps. We have observed that dissimilar distortions can be generated by the same network architecture by using different training steps or parameters. By comparing these distortions, we noticed that learning a network from previously existing databases might not be the optimal solution to our problem. On the contrary, by applying the proposed alternating training, this problem is immediately resolved: the patches reconstructed by the compression network are directly used to learn the ProxIQA network. 

\begin{figure}[!t]
  \centering
  \subfloat[Source image: Kodim01.]{\includegraphics[width=1.6in]{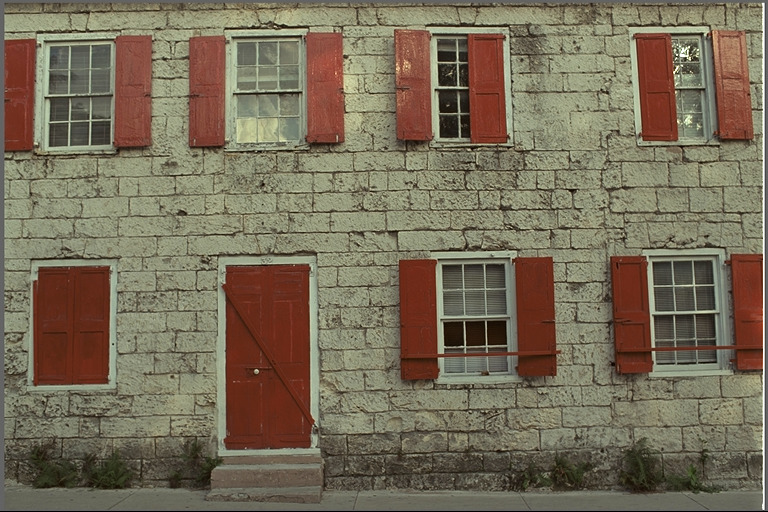}%
  \label{fig_adv_ex_src}}
  \hfil
  \subfloat[Adversarial example.]
  {\includegraphics[width=1.6in]{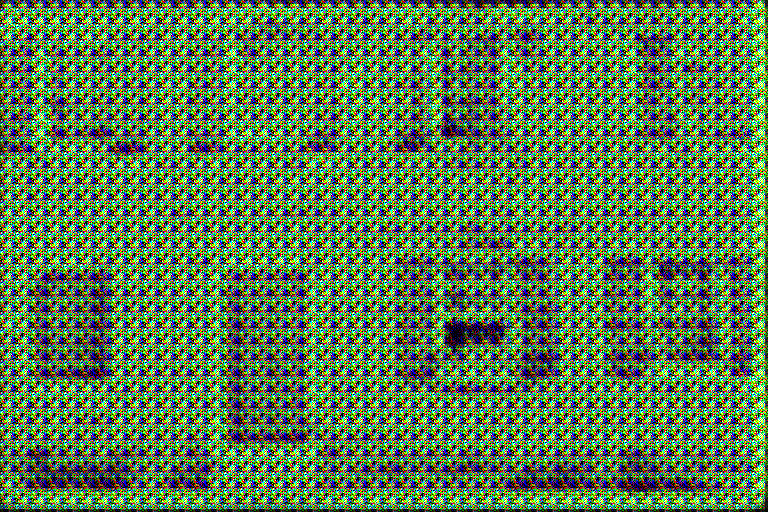}%
  \label{fig_adv_ex_rec}}
  \caption{An ``adversarial" example produced by the compression network. The $\mathrm{VMAF}$ score calculated from (a) the source image and (b) the decoded image is $5.35$ (which indicates a very poor-quality image), while the pre-trained ProxIQA network predicts a quality score of $97.74$.}
  \label{fig_adv_example}
  \end{figure}
  
  \begin{figure}[!t]
  \centering
  \subfloat[Pre-trained model.]{\includegraphics[width=2.8in]{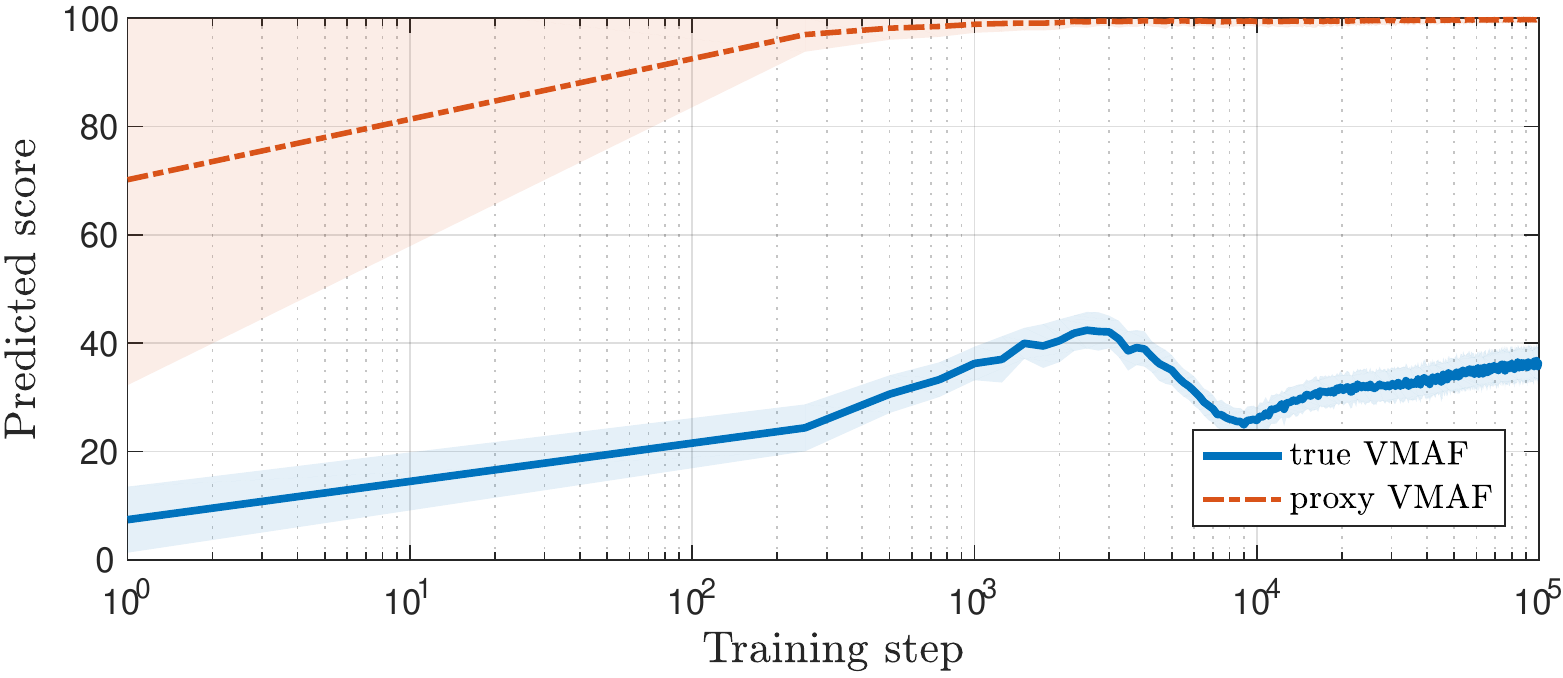}%
  \label{fig_train_plot_pre}}
  \hfil
  \subfloat[Model learned from the proposed alternating training process.]
  {\includegraphics[width=2.8in]{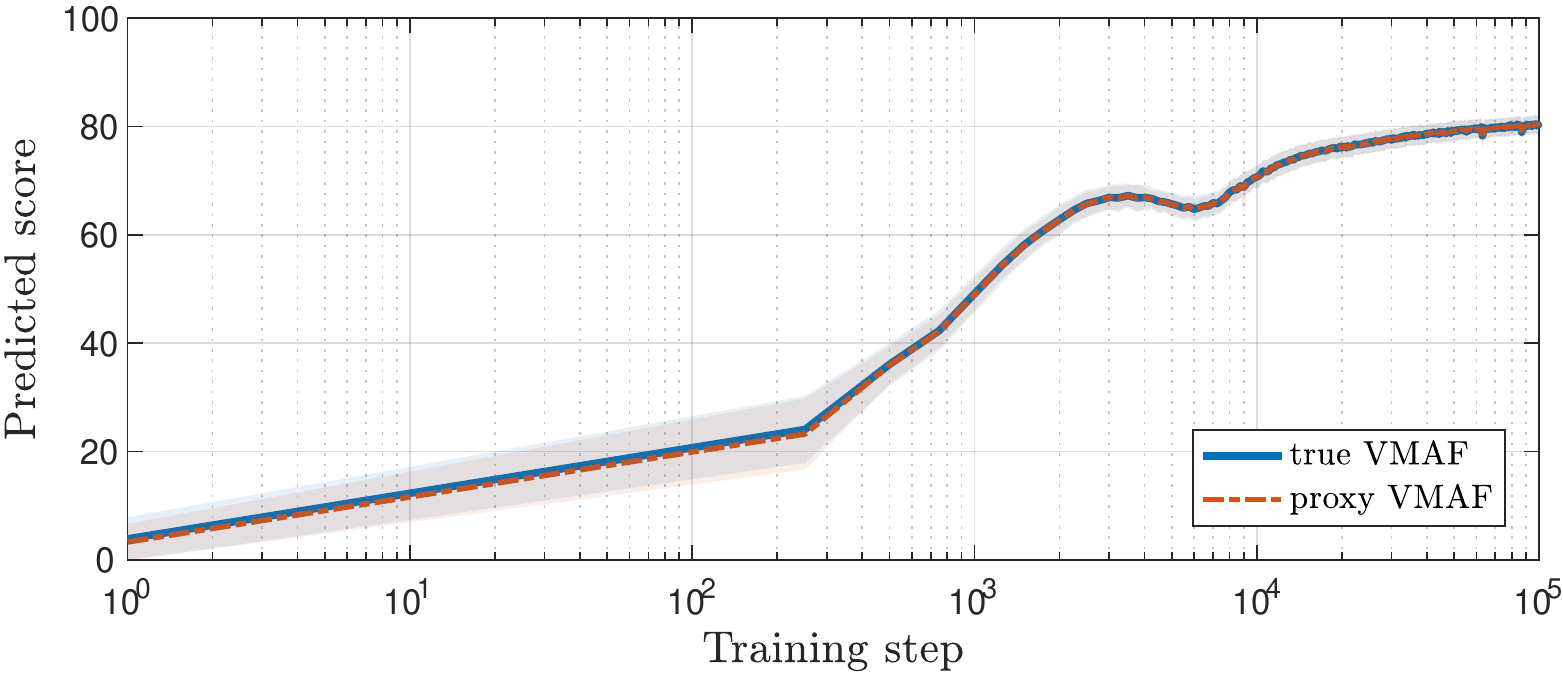}%
  \label{fig_train_plot_alt}}
  \caption{Comparison of true VMAF scores and proxy VMAF scores (quality scores predicted by ProxIQA) using two different optimization strategies during the training process. The two scores are plotted in mean values (lines) and standard deviations (shadows).}
  \label{fig_train_plot}
  \end{figure}

\subsubsection{Adversarial Examples}
We also discovered that the deep compression network often generates ``adversarial" examples when its loss layer is the output of a pre-trained network having fixed parameters. Fig.~\ref{fig_adv_example} shows such an ``adversarial" example generated by the deep compression network using a proxy network as its loss function. In this example, the proxy network was well-trained to mimic the VMAF algorithm. However, comparing Fig.~\ref{fig_adv_example}\subref{fig_adv_ex_src} with Fig.~\ref{fig_adv_example}\subref{fig_adv_ex_rec}, it is apparent that the true VMAF score and the proxy VMAF score predicted by the ProxIQA network are very different. This can be understood by considering the training of the network to be an interpolation problem, whereby the neural networks maps a test image to an accurate quality score. However, when the input is too different from the training set, the ProxIQA network may produce a poor interpolation result.

To further illustrate, Fig.~\ref{fig_train_plot} compares true VMAF scores with proxy VMAF scores. All of the scores were calculated on the reconstructed patches produced during training. Figure~\ref{fig_train_plot}\subref{fig_train_plot_pre} shows that the proxy VMAF scores quickly approached $100$, whereas much lower true VMAF scores were assigned to the patches produced by the compression model. This problem becomes significant when the previous discussed training strategy is applied. However, a straightforward way of improving the training stage is to simply feed the adversarial examples along with their objective quality scores into the ProxIQA network as additional training data. The ProxIQA network is then updated, which enables it to predict proxy quality much more accurately. As shown in Fig.~\ref{fig_train_plot}\subref{fig_train_plot_alt}, the true and proxy scores become highly coincident early in the training process. This also indicates that a CNN is able to closely approximate the responses delivered by VMAF.

\subsection{Implementation and Training Details}
The TensorFlow framework (version 1.12) was used to implement the proposed method. We use the Adam solver \cite{kingma:adam} to optimize both the ProxIQA network and the deep compression network, with $\beta_1=0.9$ and $\beta_2=0.999$. We set the initial learning rates for both networks at fixed values of $1e-4$ for the first 2M steps and $1e-5$ for an additional 100K iterations. Thus, the network was trained on 2.1M iterations of back-propagation. All of the models were trained using NVIDIA 1080-TI GPU cards.

We used a subset of the $6507$ processed images from the ImageNet database \cite{imagenet_cvpr09} as training data. As described in \cite{BalleLS16a}, small amounts of uniform noise were added to the images. The images were then down-sampled by random factors to reduce compression artifacts and high-frequency noise, and randomly cropped to a size of $256\times256$. In each mini-batch, we randomly sampled $8$ image patches from the subset. We then cropped the images to $128\times128$ patches. The source code is publicly available on https://github.com/treammm/Compression.

\section{Experiments and Analysis}
We compared the proposed perceptual optimization framework against the original MSE optimized image compression model and also against state-of-the-art image codecs. In order to experimentally evaluate the trained models, we conducted and reported the results of a quantitative evaluation, a subjective comparison, and by a runtime analysis. We first describe the experimental setups that were used, including the datasets on which the performance evaluation was conducted using standard evaluation criteria. We also performed a different series of experiments to probe the limitations of our design. In all the experiments conducted, we denote the deep image compression model \cite{BalleLS16a} as the BLS model (BLS represents the authors' first initials), which we use as a baseline for performance comparison. Also, we denote an optimized proxy model for a given IQA model $M$ using (\ref{eq:total_loss}) and (\ref{eq:metric_loss}) is denoted by $M\mathrm{_p}$.

\begin{table*}[!t]
\renewcommand{\arraystretch}{1.3}
\caption{BD-rate change (in percentage) of the optimization results of deep image compression models on the Kodak dataset, for four different IQA models. The corresponding baseline is the MSE-optimized BLS model \cite{BalleLS16a}. Smaller or negative values means better coding efficiency.}
\label{tab:kodak}
\centering
\renewcommand{\tabcolsep}{6pt} % adjust horizontal space
\begin{tabular}{l 
                r@{\hskip0.06in}r@{\hskip0.02in}r@{\hskip0.03in}r 
                r@{\hskip0.06in}r@{\hskip0.02in}r@{\hskip0.03in}r 
                r@{\hskip0.06in}r@{\hskip0.02in}r@{\hskip0.03in}r}
                % r@{\hskip0.05in}r@{\hskip0.05in}r@{\hskip0.05in}r 
                % r@{\hskip0.05in}r@{\hskip0.05in}r@{\hskip0.05in}r 
                % r@{\hskip0.05in}r@{\hskip0.05in}r@{\hskip0.05in}r}
%\toprule\toprule
\hline\hline
Optimization   & \multicolumn{4}{c}{$\text{SSIM}\mathrm{_p}$}  & \multicolumn{4}{c}{$\text{MS-SSIM}\mathrm{_p}$} & \multicolumn{4}{c}{$\text{VMAF}\mathrm{_p}$}\\
 %\cline{4-7}
BD-rate Metric &PSNR & SSIM & MS{\tiny-}SSIM & VMAF 
               &PSNR & SSIM & MS{\tiny-}SSIM & VMAF
               &PSNR & SSIM & MS{\tiny-}SSIM & VMAF\\
%\midrule
\hline
Kodim01            & \phantom{-0}14.4  & \phantom{0}-17.4 & \phantom{0}-16.3 & \phantom{00}-2.1
                   & \phantom{-0}12.6  & \phantom{00}-1.4  & \phantom{00}-3.9  & \phantom{00}-2.4 
                   & \phantom{-00}4.9   & \phantom{00}-2.9  & \phantom{00}-4.3  & \phantom{0}-25.3 \\
Kodim02            & 11.0  & -21.6 & -17.9 & 15.4
                   & 6.5   & -13.1 & -21.5 & 3.0
                   & 8.6   & -2.2  & -2.6  & -26.7 \\
Kodim03            & 13.5  & -19.5 & -16.6 & 5.8
                   & 4.3   & -14.6 & -24.1 & -10.6
                   & 7.8   & -5.2  & -7.2  & -33.2 \\
Kodim04            & 15.6  & -23.5 & -21.4 & 9.3
                   & 10.0  & -15.6 & -26.6 & 0.0 
                   & 6.3   & -6.4  & -7.4  & -26.9 \\
Kodim05            & 13.8  & -16.8 & -14.8 & 1.4
                   & 14.0  & -3.9  & -11.2 & -2.1 
                   & 3.1   & -6.1  & -6.3  & -18.1 \\
Kodim06            & 17.0  & -22.0 & -20.0 & 0.5
                   & 12.2  & -10.1 & -19.2 & -2.3
                   & 4.5   & -6.0  & -7.4  & -23.6 \\
Kodim07            & 11.6  & -14.5 & -10.9 & 8.3
                   & 7.1   & -9.9  & -21.1 & -0.6
                   & 7.0   & -10.0 & -9.7  & -24.7 \\
Kodim08            & 13.0  & -14.9 & -15.1 & -5.4
                   & 16.4  & -0.6  & -6.6  & -7.3
                   & 2.5 & -3.9 & -5.2 & -20.3 \\
Kodim09            & 16.1  & -17.6 & -14.5 & 12.3
                   & 7.6   & -12.9 & -23.8 & -1.2
                   & 6.4 & -3.1 & -4.4 & -24.9 \\
Kodim10            & 15.4  & -25.3 & -22.4 & 12.2
                   & 10.4  & -17.3 & -32.8 & 3.3
                   & 5.0 & -9.6 & -10.7 & -23.2 \\
Kodim11            & 16.6  & -26.6 & -25.3 & 12.8
                   & 14.1  & -16.1 & -25.9 & 6.9
                   & 5.1 & -7.1 & -8.5 & -20.0 \\
Kodim12            & 10.4  & -30.8 & -28.8 & 12.8
                   & 2.5   & -19.7 & -31.3 & 8.3
                   & 6.6   & -3.4  & -5.9  & -22.4 \\
Kodim13            & 19.9  & -25.3 & -24.5 & -7.1 
                   & 16.8  & -9.1  & -16.6 & -7.3
                   & 1.2   & -8.1  & -9.2  & -19.6 \\
Kodim14            & 17.5  & -22.6 & -20.3 & 8.2
                   & 13.0  & -10.5 & -18.3 & 0.0
                   & 4.0   & -8.5  & -9.2  & -20.1 \\
Kodim15            & 17.4  & -27.1 & -26.0 & 6.0
                   & 9.8   & -15.6 & -30.7 & 0.1
                   & 6.7   & -5.7  & -7.6  & -31.8 \\
Kodim16            & 14.4  & -20.4 & -15.9 & 2.5
                   & 6.3   & -13.9 & -21.3 & -3.2
                   & 6.5   & -4.6  & -6.1 & -24.9 \\
Kodim17            & 18.0  & -20.7 & -19.3 & 7.3
                   & 10.5  & -15.0 & -29.7 & 0.1
                   & 5.3   & -9.6  & -10.9 & -23.6 \\
Kodim18            & 19.7  & -20.7 & -18.3 & 12.1
                   & 17.9  & -9.2  & -17.9 & 2.4
                   & 2.7   & -9.1  & -9.8  & -22.0 \\
Kodim19            & 17.6  & -18.2 & -16.0 & 13.9
                   & 17.7  & -8.7  & -21.8 & 6.8
                   & 8.7   & -4.7  & -6.1  & -17.8 \\
Kodim20            & 19.7  & -21.3 & -22.4 & 15.3
                   & 12.8  & -10.3 & -25.3 & 0.3
                   & 5.1   & -5.4  & -8.2  & -24.2 \\
Kodim21            & 18.1  & -16.6 & -15.0 & 1.9
                   & 16.4  & -7.5  & -16.9 & -4.1
                   & 4.1   & -5.3  & -6.8  & -21.6 \\
Kodim22            & 15.0  & -23.0 & -21.2 & 11.8 
                   & 10.3  & -15.5 & -24.3 & 3.5
                   & 3.4   & -10.9 & -12.4 & -25.1 \\
Kodim23            & 14.7  & -21.6 & -17.5 & 15.5
                   & 10.5  & -16.6 & -29.6 & 4.1
                   & 7.8   & -7.5  & \phantom{0}-8.7 & -16.7 \\
Kodim24            & 21.3  & -23.4 & -21.7 & 2.0
                   & 20.6  & -11.3 & -21.9 & -1.7
                   & 2.2   & -11.4 & -12.2 & -24.0 \\
%\bottomrule\bottomrule
\hline\hline
\end{tabular}
\end{table*}

\begin{table}[!t]
  \renewcommand{\arraystretch}{1.3}
  \caption{Version of software image codecs used in the experiments}
  \label{tab:software}
  \centering
  \begin{tabular}{l l l}
  %\toprule\toprule
  \hline\hline
  Codec & Software & Version \\
  %\midrule
  \hline
  JPEG                     & JPEG XT                & Release 1.53\footnotemark[1] \\
  JPEG2000                 & Kakadu                 & Version 7.10.2\footnotemark[2] \\
  HEVC                     & HM                     & 16.9\footnotemark[3] \\
  Ball\'e\cite{BalleLS16a} & Tensorflow Compression & Release 1.0\footnotemark[4] \\
  Ball\'e\cite{balle2018variational} & Tensorflow Compression & Release 1.2\footnotemark[5] \\
  %\bottomrule\bottomrule
  \hline\hline
  \end{tabular}
  \end{table}
  
\subsection{Experimental Setup}
\subsubsection{Evaluation Datasets} To evaluate the various image codecs, we utilized the well-known Kodak dataset \cite{kodak_data} of $24$ very high quality uncompressed $768\times512$ images. This publicly available image set is commonly used to evaluate image compression algorithms and IQA models. We also used a subset of the Tecnick dataset \cite{stag_tecnick} containing $100$ images of resolution $1200\times1200$, and $223$ billboard images collected from the Netflix library having acronym NFLX, yielding images having more diverse resolutions and contents. None of the test images were included in the training sets, to avoid bias and overfitting problems.

\begin{figure*}[!ht]
	\centering
	\footnotesize
	\renewcommand{\tabcolsep}{2pt} % adjust horizontal space
	\renewcommand{\arraystretch}{1} % adjust vertical space
	\def\imgwid{0.27\textwidth}
	\def\rdhei{0.223\textwidth}
	\def\imghei{0.199\textwidth}
	\def\imgheid2{0.097\textwidth}
	
	\def\rd_shift{-0.123\textwidth}
    \def\im_shift{-0.102\textwidth}
	
	\begin{tabular}{ccccccc}
        VMAF RD-curve &
        Source Image &
        Source &
        Baseline &
        JPEG2000 &
        \VMAFp &
        HEVC
        \\\\
        %----------------- one module-------------------
        \multirow[t]{2}{*}[\rd_shift]{\includegraphics[height=\rdhei]{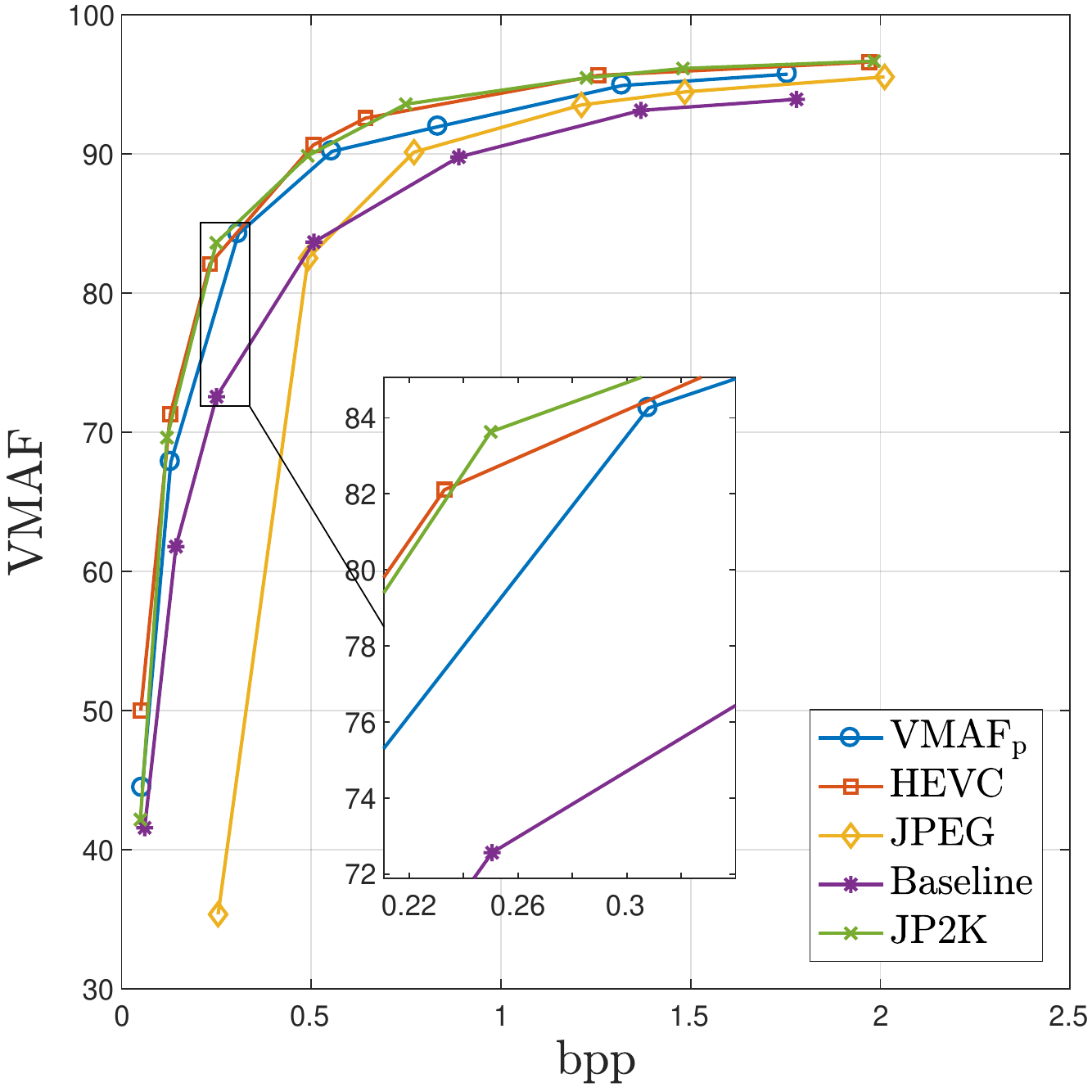}} &
        \multirow[t]{2}{*}[\im_shift]{\includegraphics[height=\imghei]{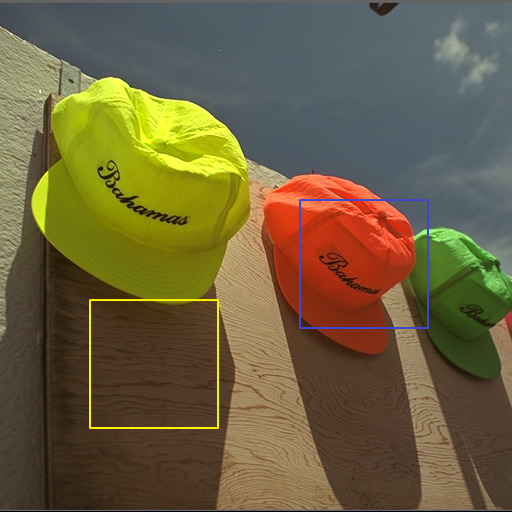}} &
        \includegraphics[height=\imgheid2]{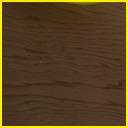} & 
        \includegraphics[height=\imgheid2]{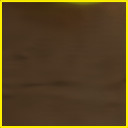} & 
        \includegraphics[height=\imgheid2]{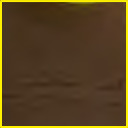} &
        \includegraphics[height=\imgheid2]{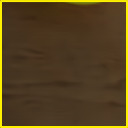} & 
        \includegraphics[height=\imgheid2]{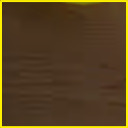} \\
		& & 
		\includegraphics[height=\imgheid2]{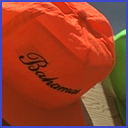} &
        \includegraphics[height=\imgheid2]{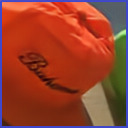} &
        \includegraphics[height=\imgheid2]{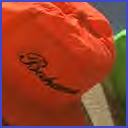} &
        \includegraphics[height=\imgheid2]{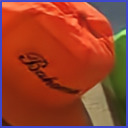} &
        \includegraphics[height=\imgheid2]{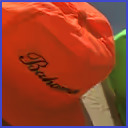} \\
                         & \multirow{2}{*}{\textbf{Kodim03}}
                         & bpp / VMAF 
                         & 0.250 / 72.56 
                         & 0.250 / 83.63 
                         & 0.307 / 84.26 
                         & 0.233 / 82.10 \\
		                 &  
                         & PSNR / SSIM 
                         & 33.02 / 0.951 
                         & 33.39 / 0.953 
                         & 33.72 / 0.963 
                         & 34.89 / 0.962 \\
        %-----------------------------------------------
        %----------------- one module-------------------
        \multirow[t]{2}{*}[\rd_shift]{\includegraphics[height=\rdhei]{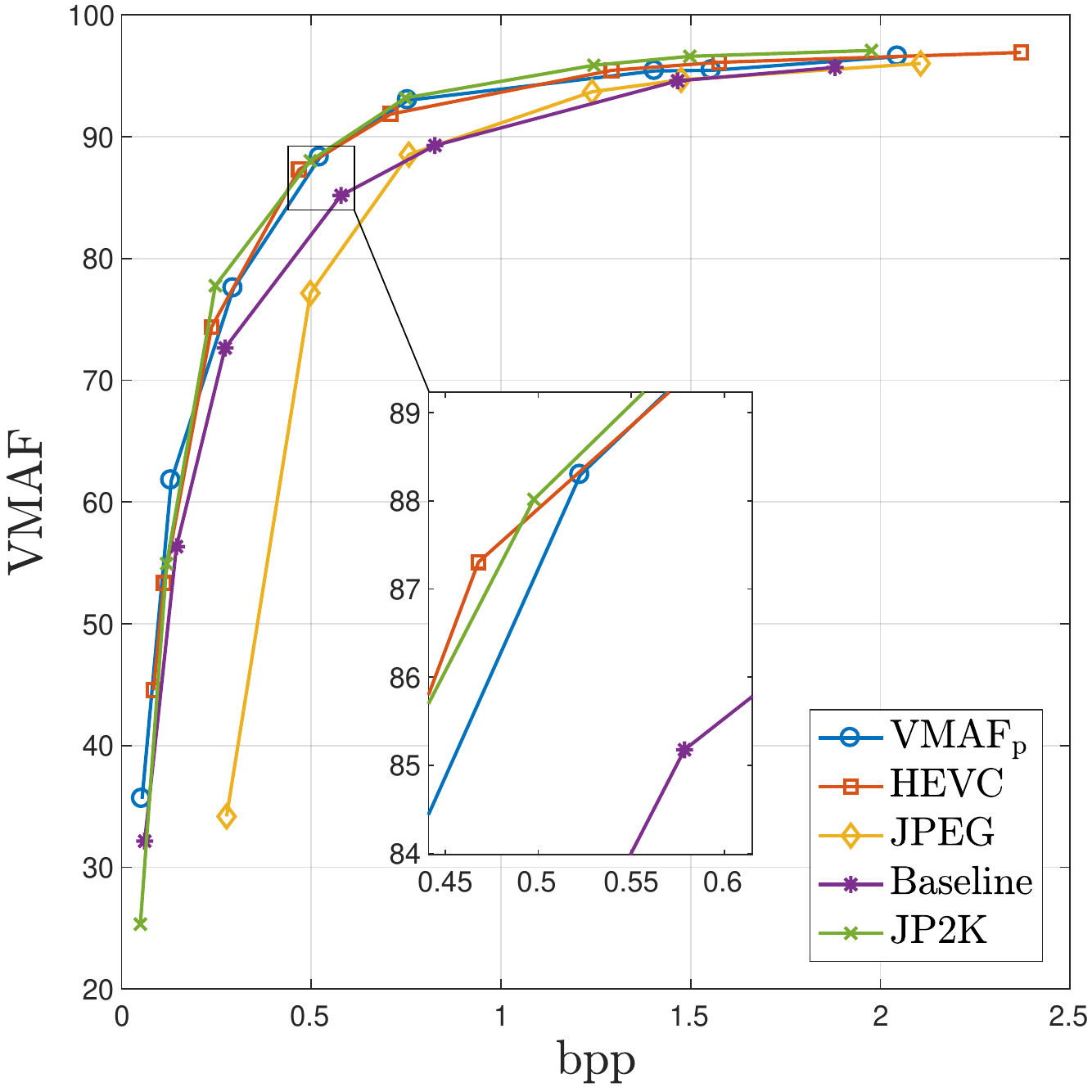}} &
        \multirow[t]{2}{*}[\im_shift]{\includegraphics[height=\imghei]{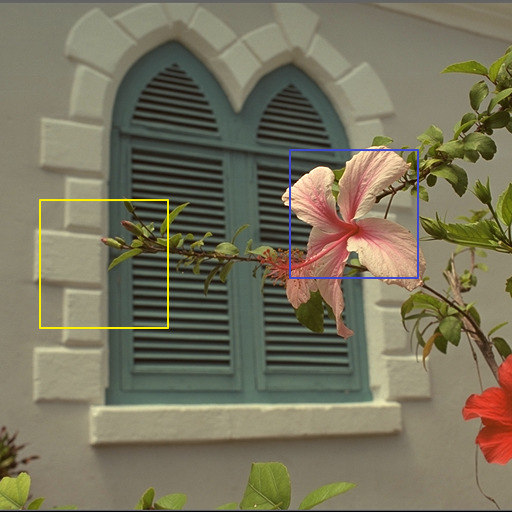}} &
        \includegraphics[height=\imgheid2]{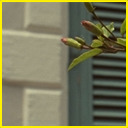} & 
        \includegraphics[height=\imgheid2]{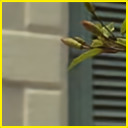} & 
        \includegraphics[height=\imgheid2]{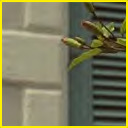} &
        \includegraphics[height=\imgheid2]{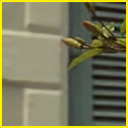} & 
        \includegraphics[height=\imgheid2]{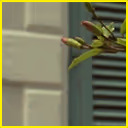} \\
		& & 
		\includegraphics[height=\imgheid2]{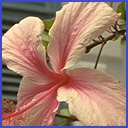} &
        \includegraphics[height=\imgheid2]{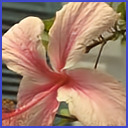} &
        \includegraphics[height=\imgheid2]{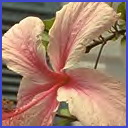} &
        \includegraphics[height=\imgheid2]{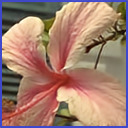} &
        \includegraphics[height=\imgheid2]{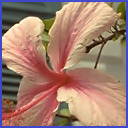} \\
                         & \multirow{2}{*}{\textbf{Kodim07}}
                         & bpp / VMAF 
                         & 0.579 / 85.17 
                         & 0.498 / 88.01 
                         & 0.522 / 88.30 
                         & 0.468 / 87.30 \\
		                 &  
                         & PSNR / SSIM 
                         & 35.71 / 0.986 
                         & 34.74 / 0.979
                         & 34.74 / 0.985 
                         & 35.77 / 0.983 \\
        %-----------------------------------------------
        %----------------- one module-------------------
        \multirow[t]{2}{*}[\rd_shift]{\includegraphics[height=\rdhei]{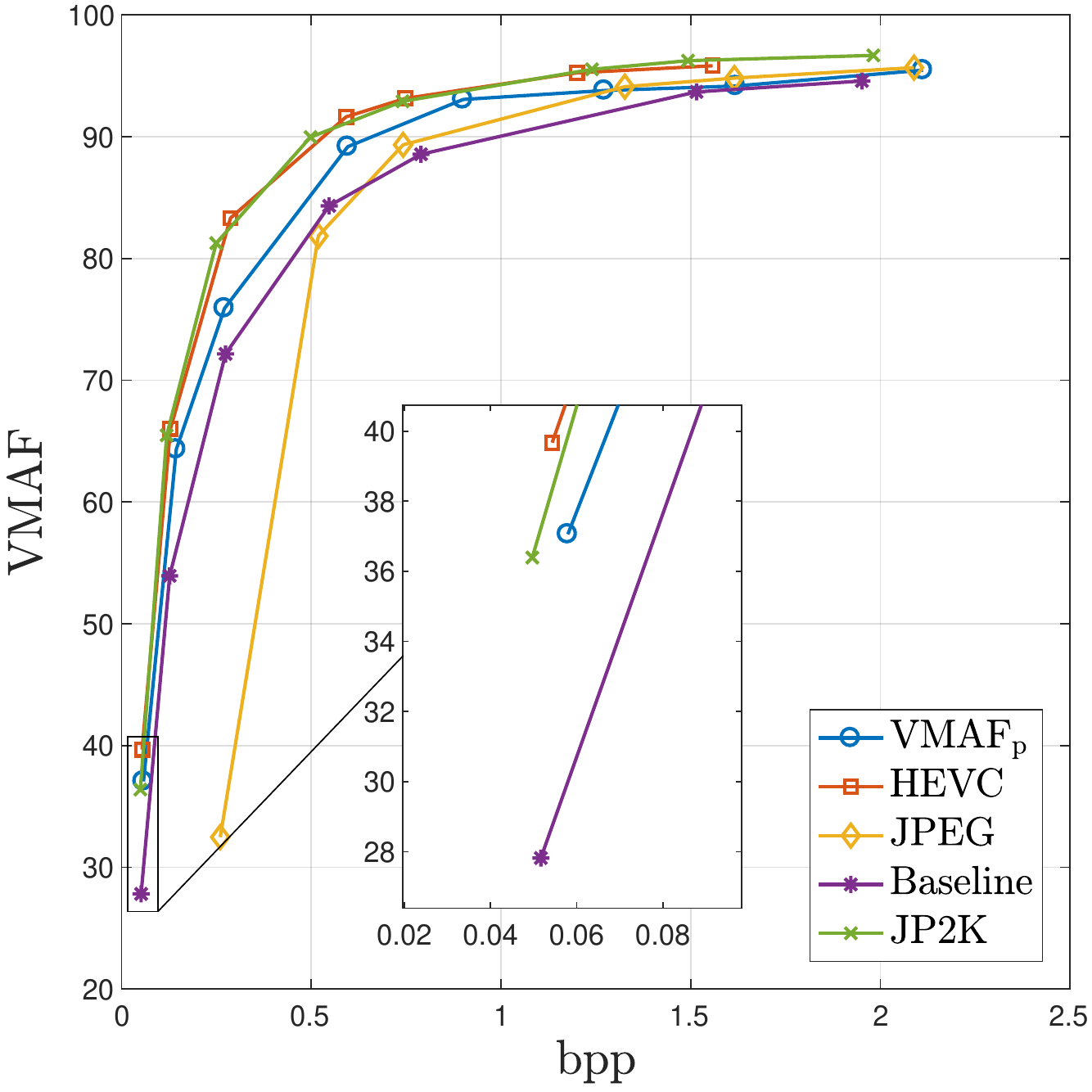}} &
        \multirow[t]{2}{*}[\im_shift]{\includegraphics[height=\imghei]{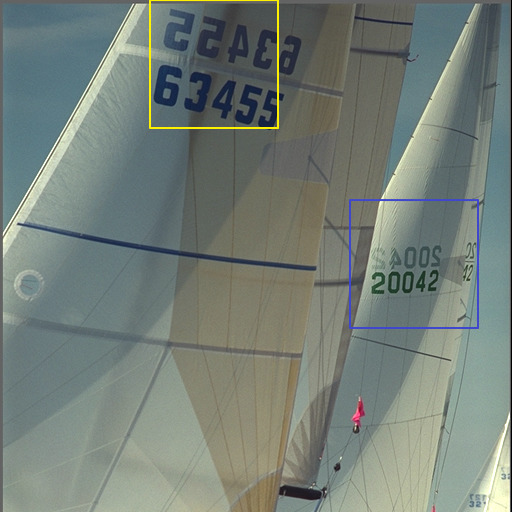}} &
        \includegraphics[height=\imgheid2]{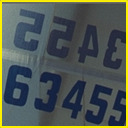} & 
        \includegraphics[height=\imgheid2]{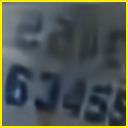} & 
        \includegraphics[height=\imgheid2]{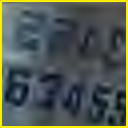} &
        \includegraphics[height=\imgheid2]{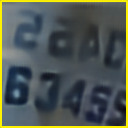} & 
        \includegraphics[height=\imgheid2]{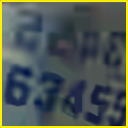} \\
		& & 
		\includegraphics[height=\imgheid2]{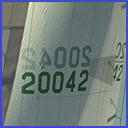} &
        \includegraphics[height=\imgheid2]{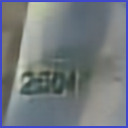} &
        \includegraphics[height=\imgheid2]{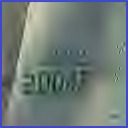} &
        \includegraphics[height=\imgheid2]{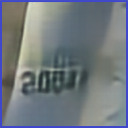} &
        \includegraphics[height=\imgheid2]{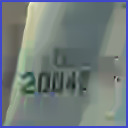} \\
                         & \multirow{2}{*}{\textbf{Kodim10}}     
                         & bpp / VMAF 
                         & 0.052 / 27.82 
                         & 0.050 / 36.39 
                         & 0.058 / 37.06 
                         & 0.054 / 39.67 \\
		                 &  
                         & PSNR / SSIM 
                         & 26.12 / 0.817 
                         & 26.53 / 0.806 
                         & 26.32 / 0.840 
                         & 27.78 / 0.843 \\
        %-----------------------------------------------
        %----------------- one module-------------------
        \multirow[t]{2}{*}[\rd_shift]{\includegraphics[height=\rdhei]{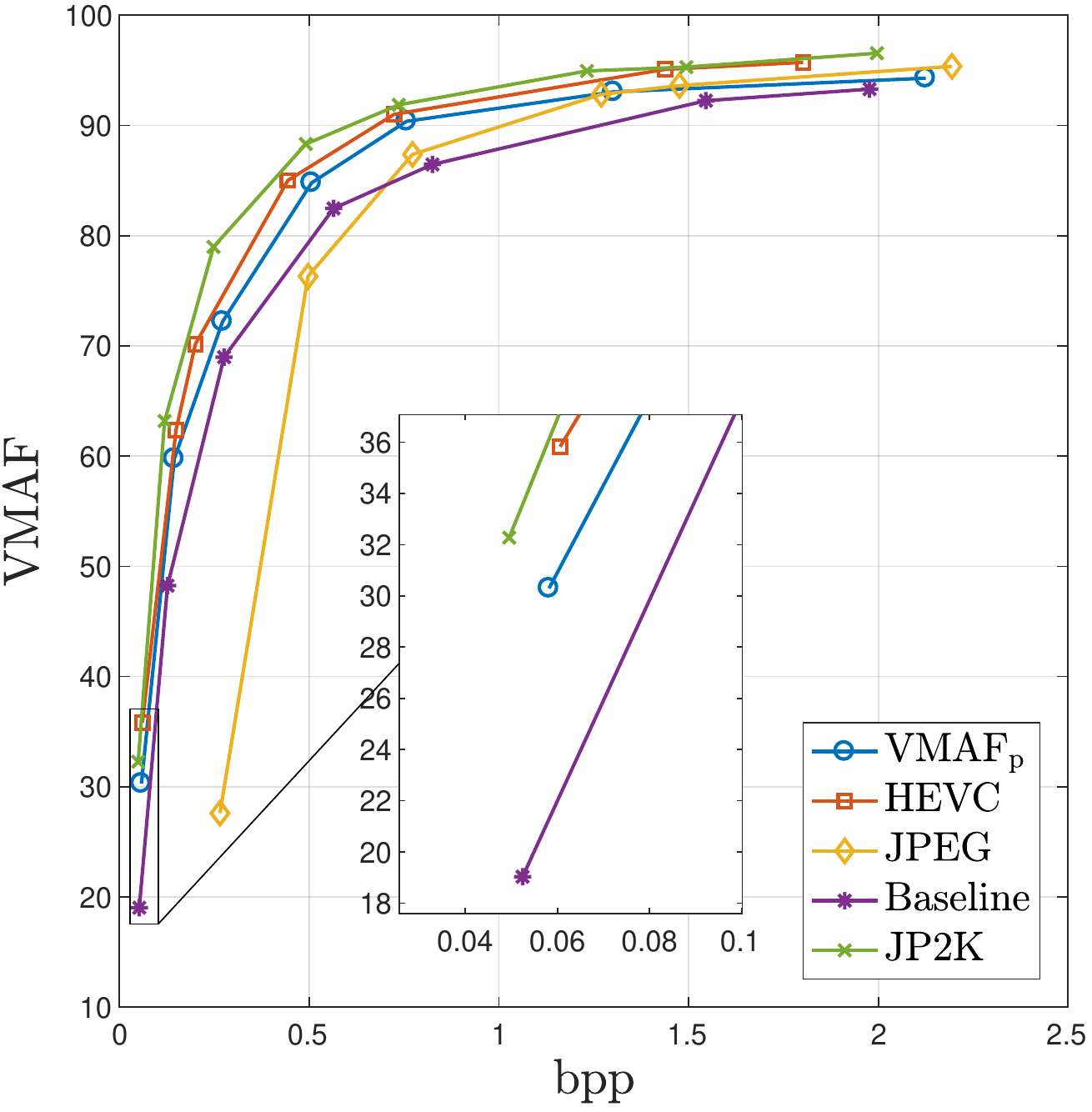}} &
        \multirow[t]{2}{*}[\im_shift]{\includegraphics[height=\imghei]{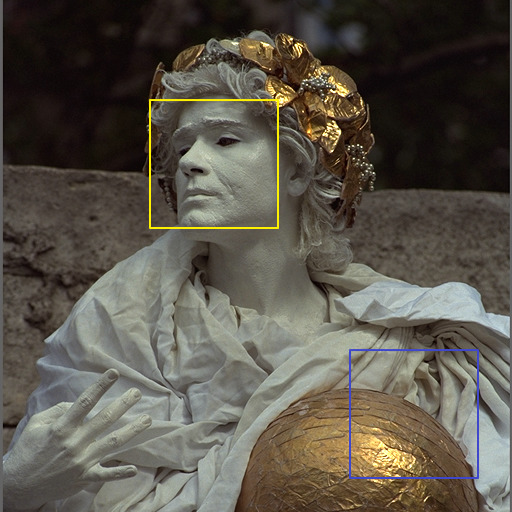}} &
        \includegraphics[height=\imgheid2]{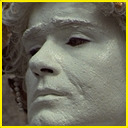} & 
        \includegraphics[height=\imgheid2]{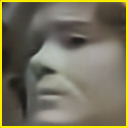} & 
        \includegraphics[height=\imgheid2]{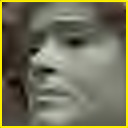} &
        \includegraphics[height=\imgheid2]{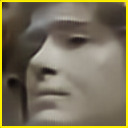} & 
        \includegraphics[height=\imgheid2]{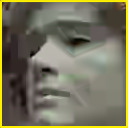} \\
		& & 
		\includegraphics[height=\imgheid2]{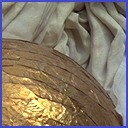} &
        \includegraphics[height=\imgheid2]{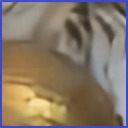} &
        \includegraphics[height=\imgheid2]{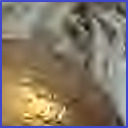} &
        \includegraphics[height=\imgheid2]{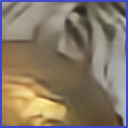} &
        \includegraphics[height=\imgheid2]{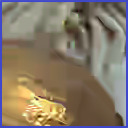} \\
                         & \multirow{2}{*}{\textbf{Kodim17}}    
                         & bpp / VMAF 
                         & 0.052 / 19.03 
                         & 0.050 / 32.27 
                         & 0.058 / 30.30 
                         & 0.061 / 35.82 \\
		                 &  
                         & PSNR / SSIM 
                         & 26.20 / 0.808 
                         & 26.03 / 0.795 
                         & 26.43 / 0.828 
                         & 27.13 / 0.821 \\
        %-----------------------------------------------
        %----------------- one module-------------------
        \multirow[t]{2}{*}[\rd_shift]{\includegraphics[height=\rdhei]{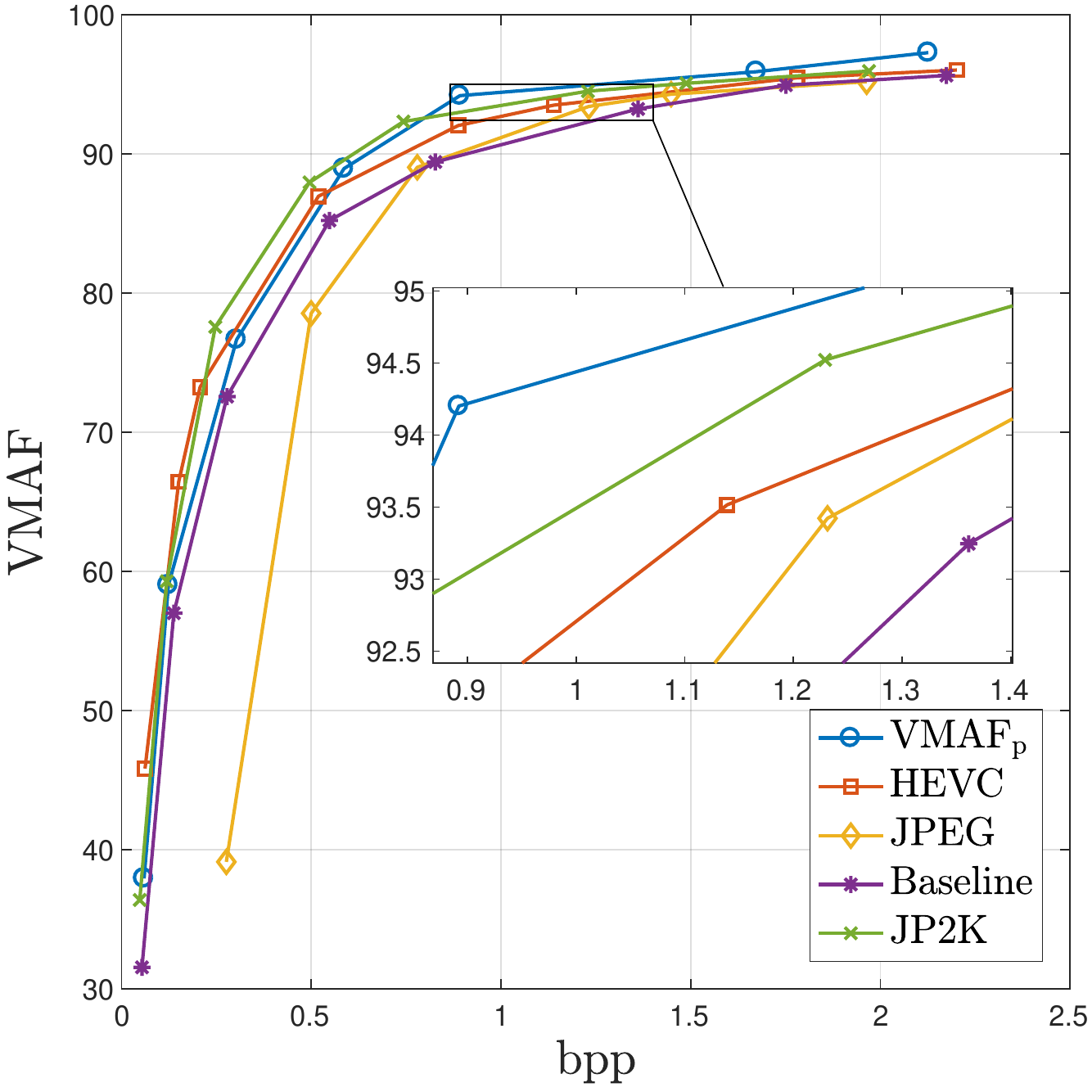}} &
        \multirow[t]{2}{*}[\im_shift]{\includegraphics[height=\imghei]{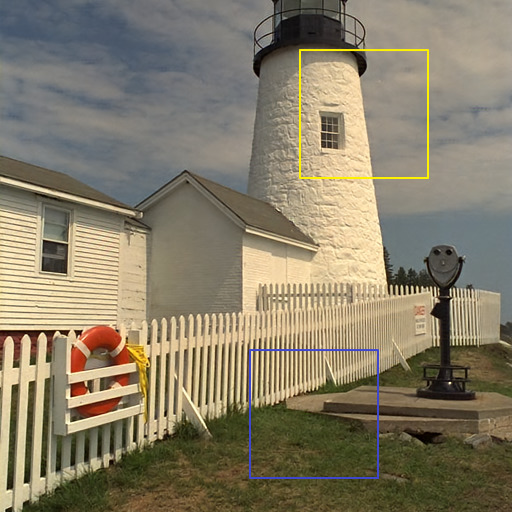}} &
        \includegraphics[height=\imgheid2]{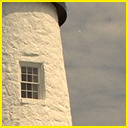} & 
        \includegraphics[height=\imgheid2]{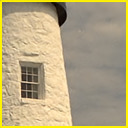} & 
        \includegraphics[height=\imgheid2]{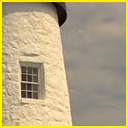} &
        \includegraphics[height=\imgheid2]{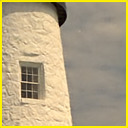} & 
        \includegraphics[height=\imgheid2]{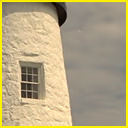} \\
		& & 
		\includegraphics[height=\imgheid2]{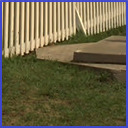} &
        \includegraphics[height=\imgheid2]{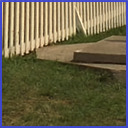} &
        \includegraphics[height=\imgheid2]{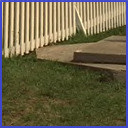} &
        \includegraphics[height=\imgheid2]{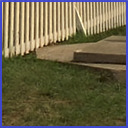} &
        \includegraphics[height=\imgheid2]{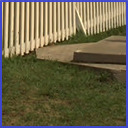} \\
                         & \multirow{2}{*}{\textbf{Kodim19}}     
                         & bpp / VMAF 
                         & 1.361 / 93.25 
                         & 1.229 / 94.52 
                         & 0.892 / 94.20 
                         & 1.139 / 93.52 \\
		                 &  
                         & PSNR / SSIM 
                         & 36.91 / 0.991 
                         & 37.52 / 0.986 
                         & 34.29 / 0.982 
                         & 37.41 / 0.986 \\
        %-----------------------------------------------
	\end{tabular}
	\caption{Visual comparison of decoded images produced by different codecs as well as the corresponding VMAF RD-curve. Image crops from left to right: ground-truth, baseline model \cite{BalleLS16a}, JPEG2000, Ball\'e et al. optimized with \VMAFp ~(denoted by \VMAFp), and HEVC. Generally, the \VMAFp-optimized BLS model achieved visual quality comparable to intra HEVC and JPEG2000. The source images were cropped to resolution $512\times512$ in the second column for display purposes.}
	\label{fig:compare_lin2}
\end{figure*}

\subsubsection{Evaluation Criteria} We measured the objective coding efficiency of each image codec using the Bjontegaard-Delta bitrate (BD-rate) \cite{BDRate01}, which quantifies differences in bitrate at a fixed distortion level relative to another reference encoder. To calculate BD-rate, we encoded the images at eight different fixed bitrates, ranging from $0.05$ bpp (bit per pixel) to $2$ bpp. The performances of all of the codecs were compared to the same baseline - the MSE-optimized BLS model. A negative number of BD-rate means the bitrate was reduced as compared with the baseline. To fairly compare deep compression models having different loss layers, we used $192$ filters at every layer, and trained all of the models using the same number of steps. To ensure reproducibility, we report the version of each software codec used in Table~\ref{tab:software}. The input image formats used were YUV444 for JPEG and JPEG2000, and both YUV420/444 for intra-coded HEVC, respectively. Lastly, the IQA algorithms that were used to evaluate the codecs were calculated using FFmpeg 4.0 with libavfilter (for PSNR) and libvmaf 0.6.1\footnotemark[6] (for SSIM, MS-SSIM, VIF, and VMAF). Specifically, the PSNR calculation in libavfilter is defined by

\begin{equation}\label{eq:psnravg}
  \mathrm{PSNR_{avg}}=\left(4\times\mathrm{PSNR_{Y}}+\mathrm{PSNR_{U}}+\mathrm{PSNR_{V}}\right)/6,
\end{equation}
which is commonly used for combining per-channel PSNRs.

\begin{table*}[!t]
\renewcommand{\arraystretch}{1.3}
\caption{Overall comparison of different codecs and results of optimized deep image compression: average change of BD-rate and standard deviation (indicated by \underline{underline}) expressed as percentage, using three different IQA models to train ProxIQA. The baseline of comparison is the MSE-optimized BLS model \cite{BalleLS16a}. Smaller or negative values indicate better coding efficiency.}
\label{tab:comparison}
\centering
\renewcommand{\tabcolsep}{3.3pt} % adjust horizontal space
\begin{tabular}{l rrrrr c rrrrr c rrrrr}
%\toprule\toprule
\hline\hline
Image Dataset  & \multicolumn{5}{c}{Kodak} &
               & \multicolumn{5}{c}{Tecnick} &
               & \multicolumn{5}{c}{NFLX} \\
                 \cline{2-6} \cline{8-12} \cline{14-18}
BD-rate Metric &PSNR & SSIM & \multicolumn{2}{r}{MS-SSIM VIF} & VMAF &
               &PSNR & SSIM & \multicolumn{2}{r}{MS-SSIM VIF} & VMAF &
               &PSNR & SSIM & \multicolumn{2}{r}{MS-SSIM VIF} & VMAF \\
%\midrule
\hline
JPEG               & 113.99 & 129.49 & 149.86 & 95.62 & 78.36 &
                   & 119.33 & 218.04 & 171.59 & 94.59 & 89.73 &
                   & 102.28 & 143.99 & 168.20 & 79.34 & 89.95 \\
                   &\ul{15.86}&\ul{21.98}&\ul{25.04}&\ul{15.20}&\ul{20.33}&
                   &\ul{31.65}&\ul{34.11}&\ul{38.88}&\ul{32.73}&\ul{24.01}&
                   &\ul{28.59}&\ul{34.94}&\ul{37.19}&\ul{25.88}&\ul{23.53}\\
JPEG2000           & -11.51 & 6.25  & -1.02 & -1.76  & -33.39 &
                   & -13.06 & -1.55 & -8.41 & -8.91  & -34.25 &
                   & -27.81 & 1.43  & -3.93 & -14.55 & -38.98 \\
                   &\ul{10.78}&\ul{14.68}&\ul{13.71}&\ul{13.19}&\ul{8.77}&
                   &\ul{22.94}&\ul{25.92}&\ul{22.89}&\ul{22.32}&\ul{18.47}&
                   &\ul{20.25}&\ul{26.84}&\ul{25.77}&\ul{23.34}&\ul{17.03}\\
HEVC               & -26.35 & -6.32 & -6.12 & -13.68& -28.23 &
                   & -28.32 & -8.97 & -11.07& -14.87& -27.65 &
                   & -49.43 & -17.12& -16.06& -29.93& -35.03 \\
                   &\ul{11.06}&\ul{18.65}&\ul{19.05}&\ul{14.14}&\ul{12.15}&
                   &\ul{22.48}&\ul{29.84}&\ul{27.56}&\ul{25.07}&\ul{24.05}&
                   &\ul{16.07}&\ul{25.63}&\ul{26.59}&\ul{21.22}&\ul{21.52}\\
HEVC$\mathrm{_{420}}$ & -27.33 & -25.98 & -24.97 & -32.73 & -42.18 &
                   & -19.48  & -28.97 & -33.95 & -40.15 & -46.67 &
                   & -37.63 & -35.41 & -33.88 & -47.51 & -50.91 \\
                   &\ul{8.56}&\ul{15.05}&\ul{15.53}&\ul{11.77}&\ul{10.32}&
                   &\ul{19.25}&\ul{22.74}&\ul{19.93}&\ul{18.05}&\ul{18.37}&
                   &\ul{18.13}&\ul{19.01}&\ul{20.08}&\ul{15.66}&\ul{16.34}\\
BLS \scriptsize MS-SSIM \cite{BalleLS16a}
                   & 107.50 & -21.89 & -27.84 & 4.46 & 11.72  &
                   & 50.31  & -15.09 & -24.17 & 6.89 & 11.25  &
                   & 67.04  & -10.58 & -24.28 & 7.12 & 28.10   \\
                   &\ul{73.42} &\ul{7.93} &\ul{5.84} &\ul{8.48}&\ul{11.98}&
                   &\ul{19.97} &\ul{9.45} &\ul{9.89} &\ul{13.49}&\ul{11.68}&
                   &\ul{21.13} &\ul{20.15} &\ul{10.85} &\ul{15.88}&\ul{13.91}\\
\textbf{BLS \scriptsize\SSIMp}
                   & 15.89  &-21.31 &-19.25 & -4.71 & 7.19  &
                   &  8.67  &-10.79 &-16.11 & -5.72 & 8.68  &
                   & 16.79  &-19.01 &-17.73 & -6.61 & 9.75 \\
                   &\ul{2.83} &\ul{3.93} &\ul{4.22} &\ul{3.44}&\ul{6.52}&
                   &\ul{6.59} &\ul{7.60} &\ul{6.13} &\ul{4.98}&\ul{6.66}&
                   &\ul{8.99} &\ul{6.82} &\ul{6.65} &\ul{6.83}&\ul{6.72}\\
\textbf{BLS \scriptsize\MSIMp}
                   & 11.67  & -11.58 & -21.77 & -3.69 & -0.17 &
                   & 4.47   & -17.40 & -23.50 & -5.00 & 0.19  &
                   & 12.28  & -11.59 & -23.53 & -5.96 & 4.34 \\
                   &\ul{4.50} &\ul{4.77}&\ul{7.17} &\ul{3.51}&\ul{4.49}&
                   &\ul{8.65} &\ul{8.55}&\ul{9.89} &\ul{5.16}&\ul{4.78}&
                   &\ul{10.40}&\ul{9.65}&\ul{10.85}&\ul{6.92}&\ul{6.44}\\
\textbf{BLS \scriptsize\VMAFp}
                   & 5.23   & -6.53 & -7.78 & -1.90 & -23.35 &
                   & 6.23   & -8.45 & -5.97 & -2.24 & -23.78 &
                   & 7.00   & -4.35 & -5.43 & -1.96 & -21.97 \\
                   &\ul{1.97}&\ul{2.57}&\ul{2.40}&\ul{1.82}&\ul{3.92}&
                   &\ul{2.91}&\ul{4.34}&\ul{4.16}&\ul{2.90}&\ul{7.16}&
                   &\ul{2.37}&\ul{5.96}&\ul{5.57}&\ul{3.66}&\ul{5.01}\\
BMSHJ \scriptsize MSE \cite{balle2018variational}
                   & -21.46 & -10.94 & -10.17 & -16.52 & -25.78 &
                   & -26.03 & -20.22 & -16.71 & -22.75 & -33.75 &
                   & -36.64 & -21.21 & -21.08 & -28.92 & -38.01 \\
                   &\ul{6.02} &\ul{6.36} &\ul{5.66} &\ul{6.22} &\ul{8.25}&
                   &\ul{10.88}&\ul{13.83}&\ul{10.68}&\ul{11.74}&\ul{12.70}&
                   &\ul{10.18}&\ul{11.32}&\ul{12.18}&\ul{12.26}&\ul{12.62}\\
BMSHJ \scriptsize MS-SSIM
                   & 93.94 & -25.52 & -34.04 & 50.77 & -11.50  &
                   & 5.55 & -32.53 & -32.55 & 31.19 & -20.90 &
                   & -3.28 & -37.82 & -42.52 & 27.34 & -25.74 \\
                   &\ul{77.19} &\ul{21.98} &\ul{25.04} &\ul{15.20}&\ul{20.33}&
                   &\ul{12.77} &\ul{34.11} &\ul{38.83} &\ul{32.73}&\ul{24.01}&
                   &\ul{18.47} &\ul{14.43} &\ul{13.93} &\ul{34.49}&\ul{16.75}\\
\textbf{BMSHJ \scriptsize\VMAFp}
                   & -15.90 & -13.57 & -13.17 & -17.02 & -47.11 &
                   & -19.64 & -23.14 & -16.73 & -23.63 & -53.18 &
                   & -29.96 & -18.87 & -19.29 & -29.00 & -56.06 \\
                   &\ul{5.34} &\ul{4.22} &\ul{3.70} &\ul{4.86} &\ul{7.72}&
                   &\ul{10.35}&\ul{10.29}&\ul{10.82}&\ul{9.69} &\ul{11.63}&
                   &\ul{10.34}&\ul{11.01}&\ul{10.41}&\ul{10.87}&\ul{10.66}\\
%\bottomrule\bottomrule
\hline\hline
\end{tabular}
\end{table*}

\footnotetext[1]{https://jpeg.org/downloads/jpegxt/reference1367abcd89.zip}
\footnotetext[2]{https://kakadusoftware.com/downloads/}
\footnotetext[3]{https://hevc.hhi.fraunhofer.de/svn/svn\_HEVCSoftware/tags/HM-16.9/}
\footnotetext[4]{https://github.com/tensorflow/compression/releases/tag/v1.0}
\footnotetext[5]{https://github.com/tensorflow/compression/releases/tag/v1.2}
\footnotetext[6]{https://github.com/Netflix/vmaf}

\subsection{Experimental Results on Kodak Dataset} The results on the images in the Kodak dataset are given Table~\ref{tab:kodak}. The distortion levels that were used for BD-rate calculation were quantified using PSNR, SSIM, MS-SSIM, and VMAF. It may be observed that the RD performances that were measured using PSNR became worse on models optimized under other IQA models. This is not surprising since MSE, which is used by the baseline model, is the optimal loss function for PSNR. In fact, this phenomenon may also be related to the distortion-perception tradeoff described in \cite{Blau2018}: given a fixed rate, the improvement of perceptual quality must be balanced against the loss in distortion. The second thing to notice is the performance attained by the \SSIMp~and \MSIMp~optimizations. Both the SSIM and MS-SSIM BD-rates were improved significantly, as expected. Yet, the VMAF BD-rate performed slightly worse than the original MSE optimization, indicating there exists disagreements between VMAF and SSIM/MS-SSIM. Similar trends can also be observed from the MS-SSIM optimization given in Table \ref{tab:comparison}.

It may also be noted that, unlike using other IQA models as targets of the proposed optimization, \VMAFp~optimization delivers coding gain with respect to all of the BD-rate measurements, except the PSNR BD-rate. Since PSNR is known not to correlate as well with human subjective quality judgments as do modern VQA models, the improvements on the PSNR BD-rate may not always mean subjective improvements. This suggests that VMAF is a good optimization target.

\subsection{Comparison with State-of-the-art Codecs}
Table~\ref{tab:comparison} tabulates the percent change in BD-rate relative to the BLS baseline, with respect to different quality models. We comprehensively evaluated perceptual deep compression using different perceptual optimization protocols (highlighted in boldface), against three conventional image codecs: JPEG, JPEG2000, and intra-coded HEVC main-RExt (Format Range Extension) profile. Extensive experiments were carried out on the three aforementioned datasets, using three perceptual IQA models as optimization targets. In addition to the BLS model, we also deployed the proposed \VMAFp~optimization framework on a more sophisticated deep compression model BMSHJ \cite{balle2018variational}, representing the authors' first initials, to test its generality. We report the BD-rate changes obtained, averaged over all the images in each dataset. The reported standard deviation was also calculated from all the BD-rate changes within a test dataset. Similar results were obtained on the Tecnick and NFLX datasets, as shown in Tables~\ref{tab:kodak} and \ref{tab:comparison}. These results show that our optimization approach is able to successfully optimize a deep image compression model relative to different IQA algorithms. Indeed, significant BD-rate reductions were obtained in many cases. We were able to demonstrate that the BLS model optimized by $M_\mathrm{p}$ achieved similar or better results than HEVC, when the result was measured by the perceptual quality model $M$. It should be noted that the quality measurements made by SSIM, MS-SSIM, VIF, and VMAF only consider distortions of the luma channel, hence, the remaining scope for improvement by also accounting for the quality contributions of the chroma channels. Nevertheless, the \VMAFp-optimized BMSHJ model still outperformed HEVC$\mathrm{_{420}}$, in the sense of BD-rate as measured by VMAF.

In addition to the quantitative results, we also visually compared the decoded images. Fig.~\ref{fig:compare_lin2} plots VMAF Rate-distortion (RD) curves for several images. A subset of the images corresponding to the RD points obtained by the various codecs are also shown. As a basic test, we subjectively compare results yielding similar bitrates but different objective quality scores. The images Kodim10 and Kodim17 were subject to extreme compression at bitrates around $0.05$ bpp. In these cases, the \VMAFp-optimized model significantly outperformed the MSE-optimized baseline model, delivering performance comparable to HEVC and JPEG2000 with respect to VMAF score and subjective quality. At high bitrates, the distinctions between the codecs becomes subtle. Therefore, we isolated these RD points associated with similar VMAF scores and compared bitrate consumption. Generally, the $\text{VMAF}\mathrm{_p}$-optimized model yielded comparable subjective and objective (VMAF) quality as the baseline $\mathrm{MSE}$-optimized model, while consuming significantly fewer bits. The encoding results on Kodim19 in Fig.~\ref{fig:compare_lin2} using the $\text{VMAF}\mathrm{_p}$-optimized model yielded similar VMAF scores as the other codecs, while consuming only $34.4\%$, $27.4\%$ and $21.7\%$ fewer bits than the Baseline, JPEG2000, and HEVC, respectively.

\begin{figure}[!t]
  \centering
  \subfloat[Kodim03 ($\text{BD-rate}=-9.06\%$)]{\includegraphics[width=1.74in]{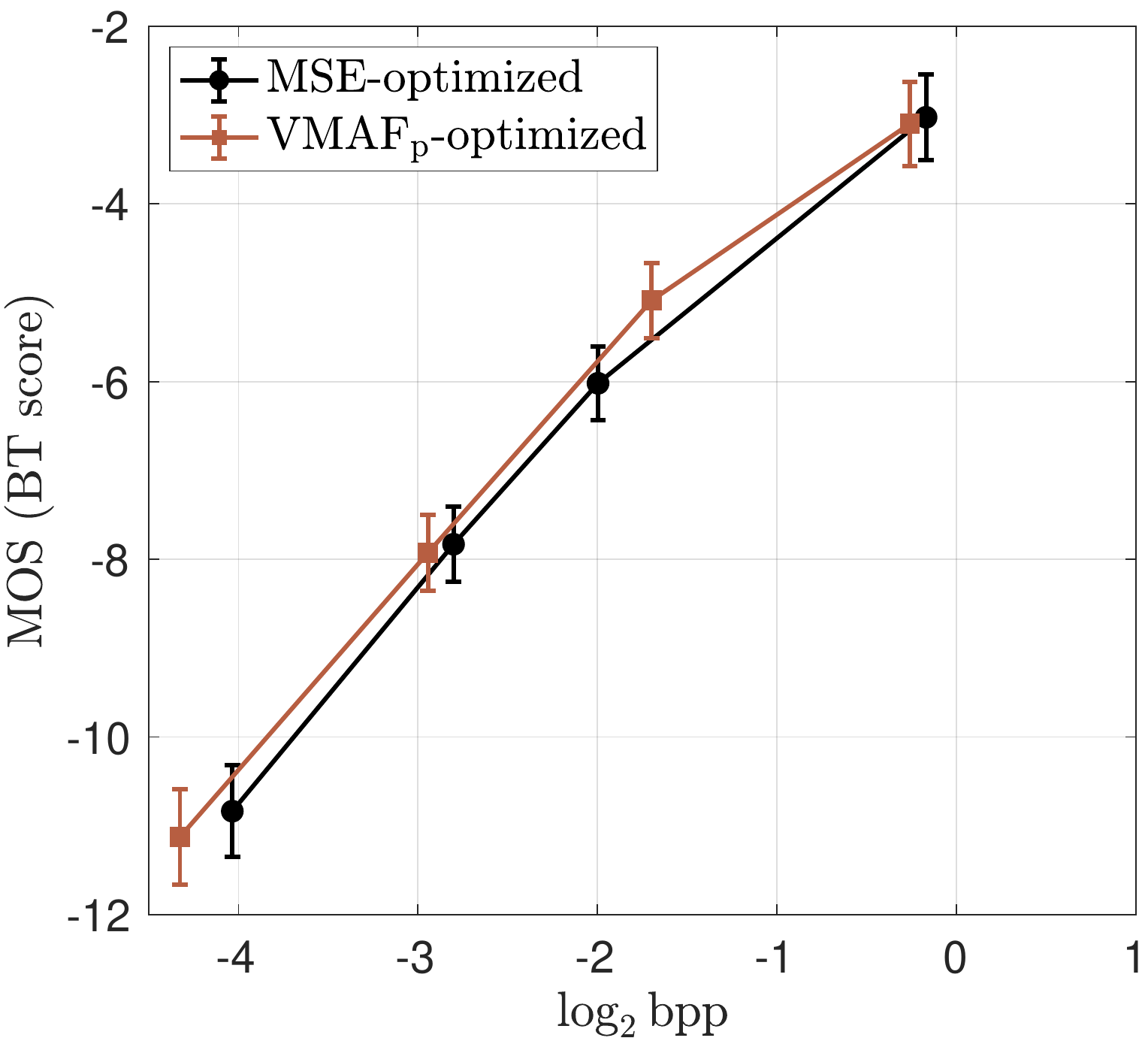}%
  \label{fig_rd_mos_k03}}
  \hfil
  \subfloat[Kodim07 ($\text{BD-rate}=-32.32\%$)]{\includegraphics[width=1.74in]{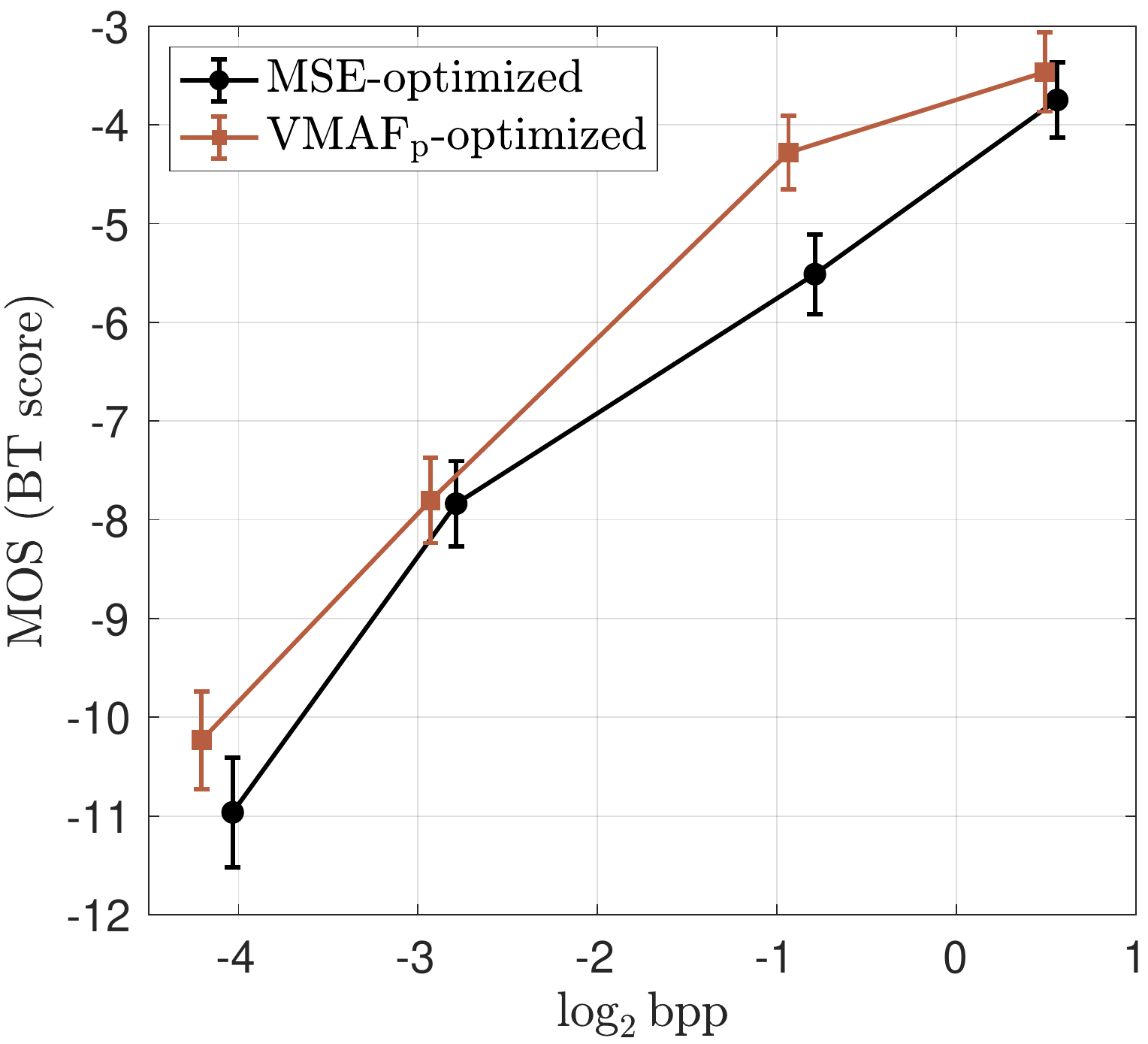}%
  \label{fig_rd_mos_k21}}
  \caption{Subjective performance comparison of rate-distortion curves (with $95\%$ confidence intervals) for two different optimization protocols on two exemplar Kodak images. The y-axis denotes the MOS scores estimated by the BT model.}
  \label{fig_rd_mos_comp}
\end{figure}

\subsection{Subjective Quality Study}
To better understand perceptual preferences on compressed images and the effects of perceptual optimization using ProxIQA, we conducted a human subject study where we compared deep compression engines using different optimization targets. Specifically, we randomly selected $10$ image contents from the Kodak dataset. Each content was encoded at four different bitrates by \VMAFp-optimized and MSE-optimized BLS models, resulting in eight different compression-distorted versions of that content. Since compressed images obtained using different algorithms often have subtle perceptual differences at similar bitrates, it is easier for humans to make comparisons between simultaneously displayed images. Therefore, we adopted the paired comparison method instead of the Absolute Category Rating (ACR) protocol in our experimental design. We designed a web-based user interface to carry out the human study, whereby participants could easily compare pairs of images and render relative quality decisions. On each image pair, the subject was asked to select the one they preferred. We recruited $41$ volunteer participants, of whom about $70\%$ were knowledgeable about image/video processing, while the others were naive. We asked each participant to compare $280$ pairs of images, which required about $30$ minutes. To obtain human opinion scores, we deployed the Bradley-Terry (BT) model \cite{btscore1952} to estimate the Mean Opinion Score (MOS).

Given the collected subjective quality score and bitrate for each distorted picture, we compared the RD performances of the two optimization methods. On the $10$ selected contents, the \VMAFp-optimized model was able to achieve reduction of MOS BD-rate ranging from $9\%$ to $32.3\%$ relative to MSE optimization, with an average reduction of $17.7\%$. Fig. \ref{fig_rd_mos_comp} shows the MOS RD curves of two contents, one of which was the most improved using \VMAFp~optimization, while the other was the least improved. It may be observed that the RD curves resulting from \VMAFp~optimization allow for reduced bitrates at similar levels of subjective quality, as compared to those produced by MSE optimization. For example, Fig. \ref{fig_rd_mos_comp}(b) shows that \VMAFp~delivers better predicted visual quality, with statistical significance, while consuming fewer bits at around $0.5$ bpp. These results show that the proposed \VMAFp~optimization yields favorable compression results as compared to MSE optimization.

\begin{table}[!t]
  \renewcommand{\arraystretch}{1.3}
  \caption{Average change in BD-rate (in percentage) of different SSIM-driven optimization results on Kodak dataset.}
  \label{tab:compare_ssim}
  \centering
  \begin{tabular}{l rrr}
  %\toprule\toprule
  \hline\hline
      & SSIM (\ref{eq:ssim_loss})
      & \SSIMp~(\ref{eq:total_loss})(\ref{eq:metric_loss})
      & SSIM+$\mathcal{L}_{d}$ (\ref{eq:total_ssim_l1_loss}) \\
  %\midrule
  \hline
  PSNR BD-Rate    & 133.79 & 15.56  & 15.43\\
  SSIM BD-Rate    & -29.41 & -19.04 & -22.39\\
  MS-SSIM BD-Rate & -11.74 & -17.29 & -19.69\\
  VMAF BD-Rate    & 40.09  & 7.63   & 12.03\\
  \hline\hline
  \end{tabular}
  \end{table}

\subsection{Limitations}
When measuring RD performance using SSIM, optimizing a model using SSIM should approach the theoretical upper bound of SSIM-measured RD performance. Accordingly, we investigated the performance of our proposed framework by comparing SSIM and \SSIMp ~optimization of the BLS model. The results of comparison are presented in Table~\ref{tab:compare_ssim}, indicating a $10\%$ SSIM BD-rate performance gap between the two optimization approaches. A noticeable contributor to this performance drop is the pixel loss $\mathcal{L}_d$ in (\ref{eq:total_loss}). To validate this assumption, we conducted an ablation study to pinpoint the cause of this gap. We did this by fixing $\mathcal{L}_p=\mathcal{L}_{\text{SSIM}}$ for the SSIM optimization. The loss function is then
\begin{equation}\label{eq:total_ssim_l1_loss}
  \mathcal{L}
  =\lambda\left[ \alpha\mathcal{L}_{\text{SSIM}} + \left( 1-\alpha \right) \mathcal{L}_{d} \right] + \mathcal{L}_{r},
\end{equation}
where
\begin{equation}\label{eq:ssim_loss}
  \mathcal{L}_{\text{SSIM}} = 1-\text{SSIM}\left( \mathbf{x},\hat{\mathbf{x}} \right),
\end{equation}
and $\alpha=3e-3$ is the same as in (\ref{eq:total_loss}). The SSIM BD-rate that resulted from optimization of (\ref{eq:total_ssim_l1_loss}) is given in the fourth column of Table~\ref{tab:compare_ssim}. It may be observed that the RD performance becomes very close to that of \SSIMp ~optimization, which confirms that the pixel loss is the main contributor to the performance loss.

\begin{figure}[!t]
  \centering
  \subfloat[Kodim01.]{\includegraphics[width=1.7in]{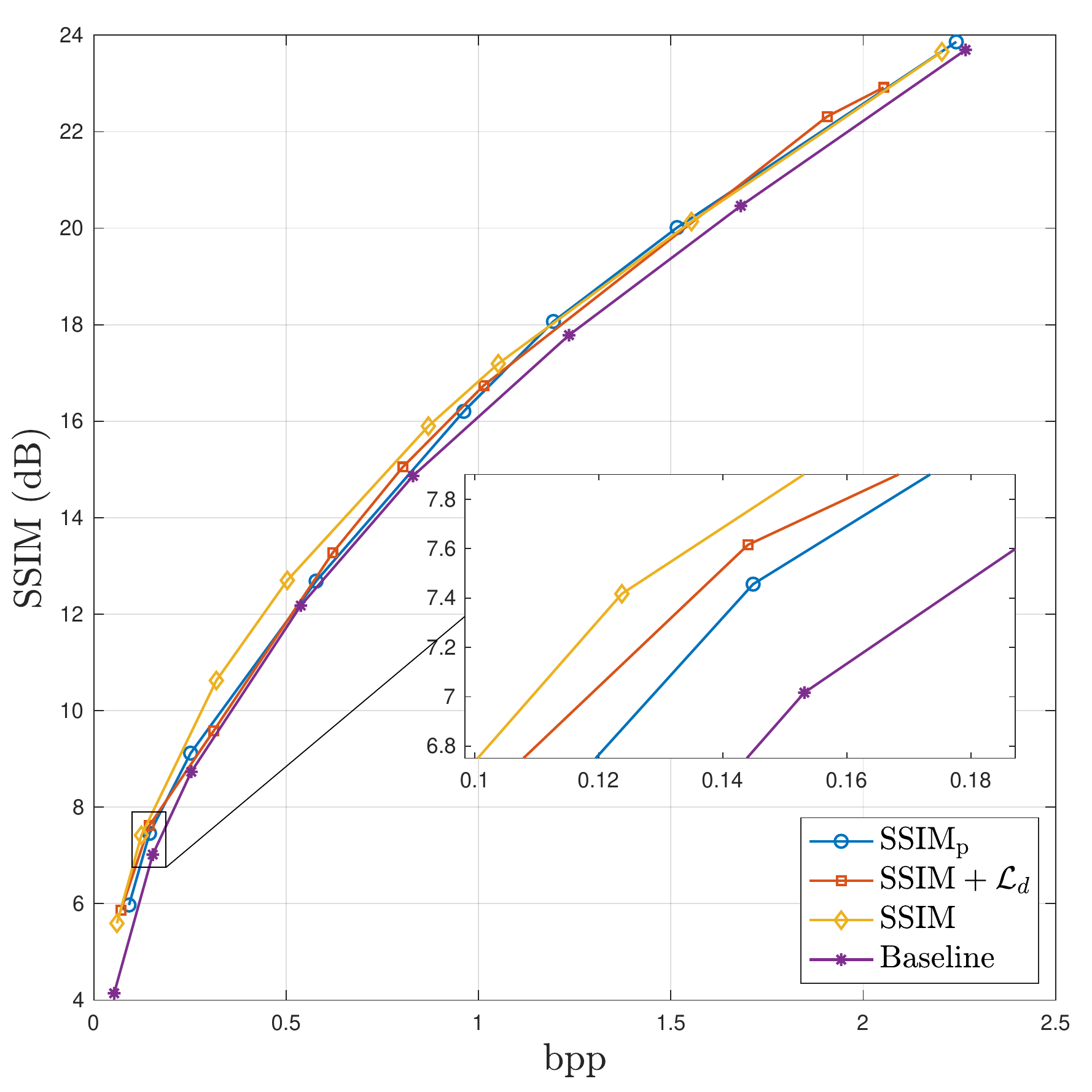}%
  \label{fig_rd_ssim_k01}}
  \hfil
  \subfloat[Kodim21.]{\includegraphics[width=1.7in]{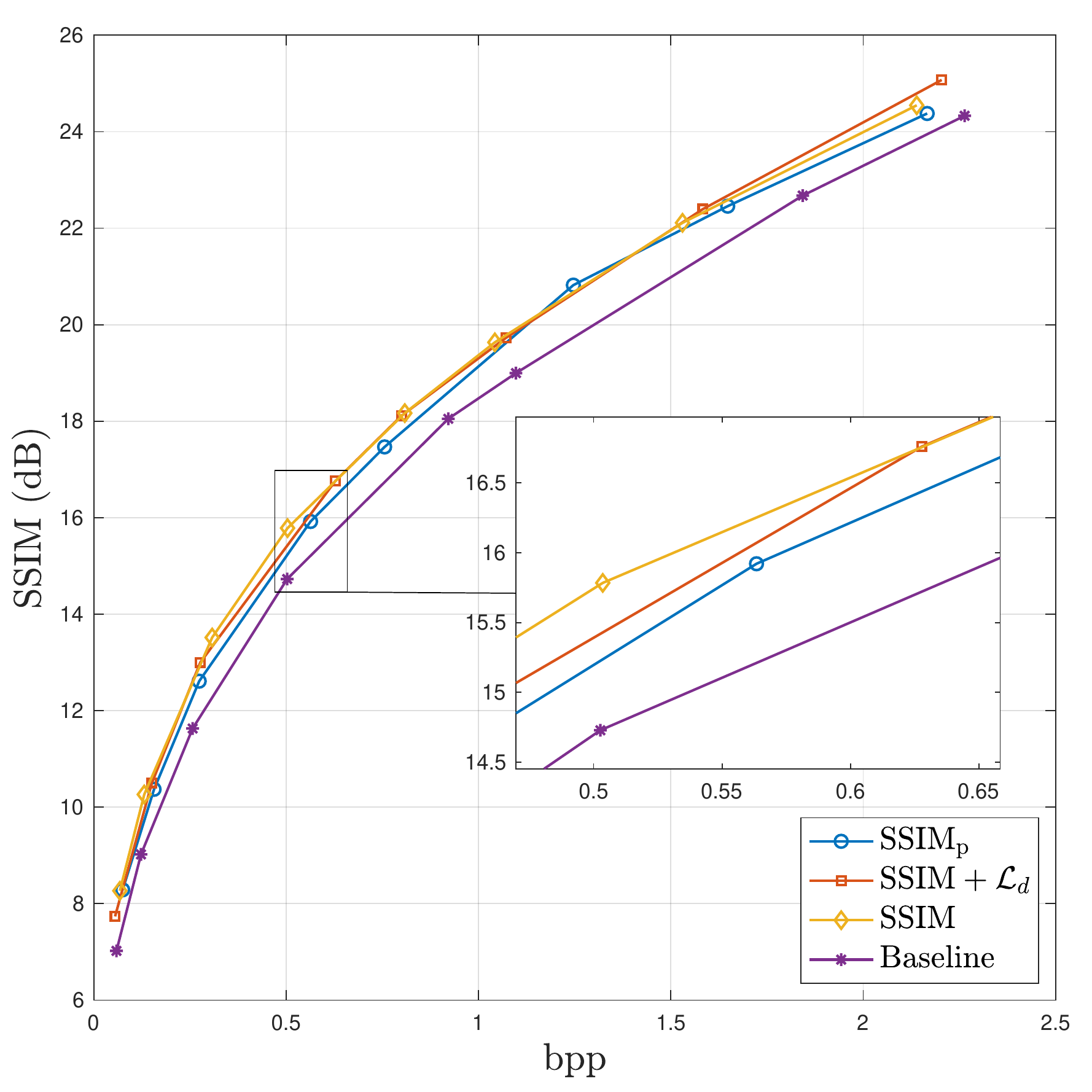}%
  \label{fig_rd_ssim_k21}}
  \caption{Rate-distortion curves for different SSIM-oriented optimization protocols on two Kodak images. The baseline curve denotes the MSE-optimized BLS model \cite{BalleLS16a}.}
  \label{fig_rd_ssim_comp}
\end{figure}

\begin{figure}[!t]
	\centering
	\footnotesize
	\renewcommand{\tabcolsep}{1pt} % adjust horizontal space
	\renewcommand{\arraystretch}{1} % adjust vertical space
	\def\imgheid2{0.118\textwidth}

	\begin{tabular}{cccc}
        %----------------- one module-------------------
        Source & 
        SSIM (\ref{eq:ssim_loss})& 
        \SSIMp~(\ref{eq:total_loss})(\ref{eq:metric_loss})& 
        SSIM+$\mathcal{L}_{d}$ (\ref{eq:total_ssim_l1_loss}) \\
        \includegraphics[height=\imgheid2]{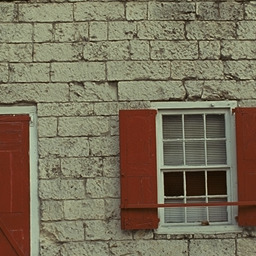} & 
        \includegraphics[height=\imgheid2]{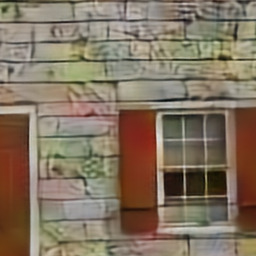} &
        \includegraphics[height=\imgheid2]{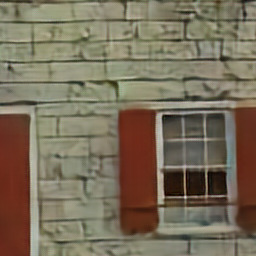} & 
        \includegraphics[height=\imgheid2]{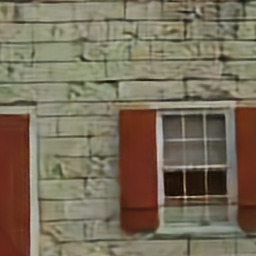} \\
        bpp / SSIM & 
        0.124 / 0.819 & 
        0.145 / 0.820 & 
        0.144 / 0.827 \\
		\includegraphics[height=\imgheid2]{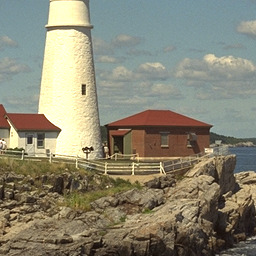} &
        \includegraphics[height=\imgheid2]{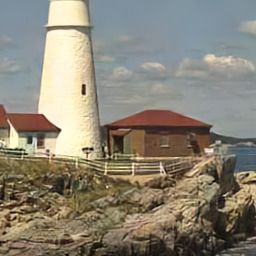} &
        \includegraphics[height=\imgheid2]{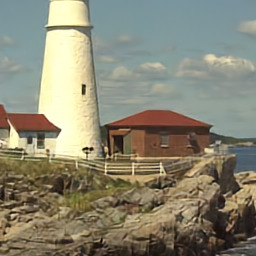} &
        \includegraphics[height=\imgheid2]{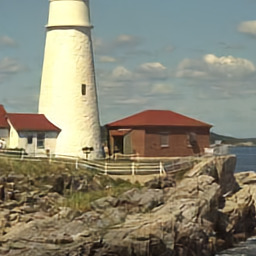} \\
        bpp / SSIM & 
        0.504 / 0.974 & 
        0.563 / 0.974 & 
        0.628 / 0.979 \\
        %-----------------------------------------------

	\end{tabular}
	\caption{Visual comparison of model behavior among different SSIM-optimized models. First row: Kodim01 encoded at $0.12$ bpp. Second row: Kodim21 encoded at $0.5$ bpp.}
	\label{fig:compare_ssim}
\end{figure}

Moreover, Fig.~\ref{fig:compare_ssim} shows that similar visual results are obtained using \SSIMp~optimization and the optimization described in (\ref{eq:total_ssim_l1_loss}), even at heavy compression levels. A close examination shows that true SSIM optimization nicely preserves high-frequency details but loses chromatic fidelity. The RD-curves in Fig.~\ref{fig_rd_ssim_comp} further confirm the similar behavior of \SSIMp ~optimization and the optimization of (\ref{eq:total_ssim_l1_loss}). We present SSIM in decibels for readability, since small quantitative differences in SSIM may be associated with large visual differences.

All the models described in this subsection were trained using one million steps and a constant learning rate. Thus, the performance results of \SSIMp~differ slightly from the results reported in Tables~\ref{tab:kodak} and \ref{tab:comparison}.

\subsection{Study of Training Steps}
\begin{figure}[!t]
  \centering
  \includegraphics[width=3in]{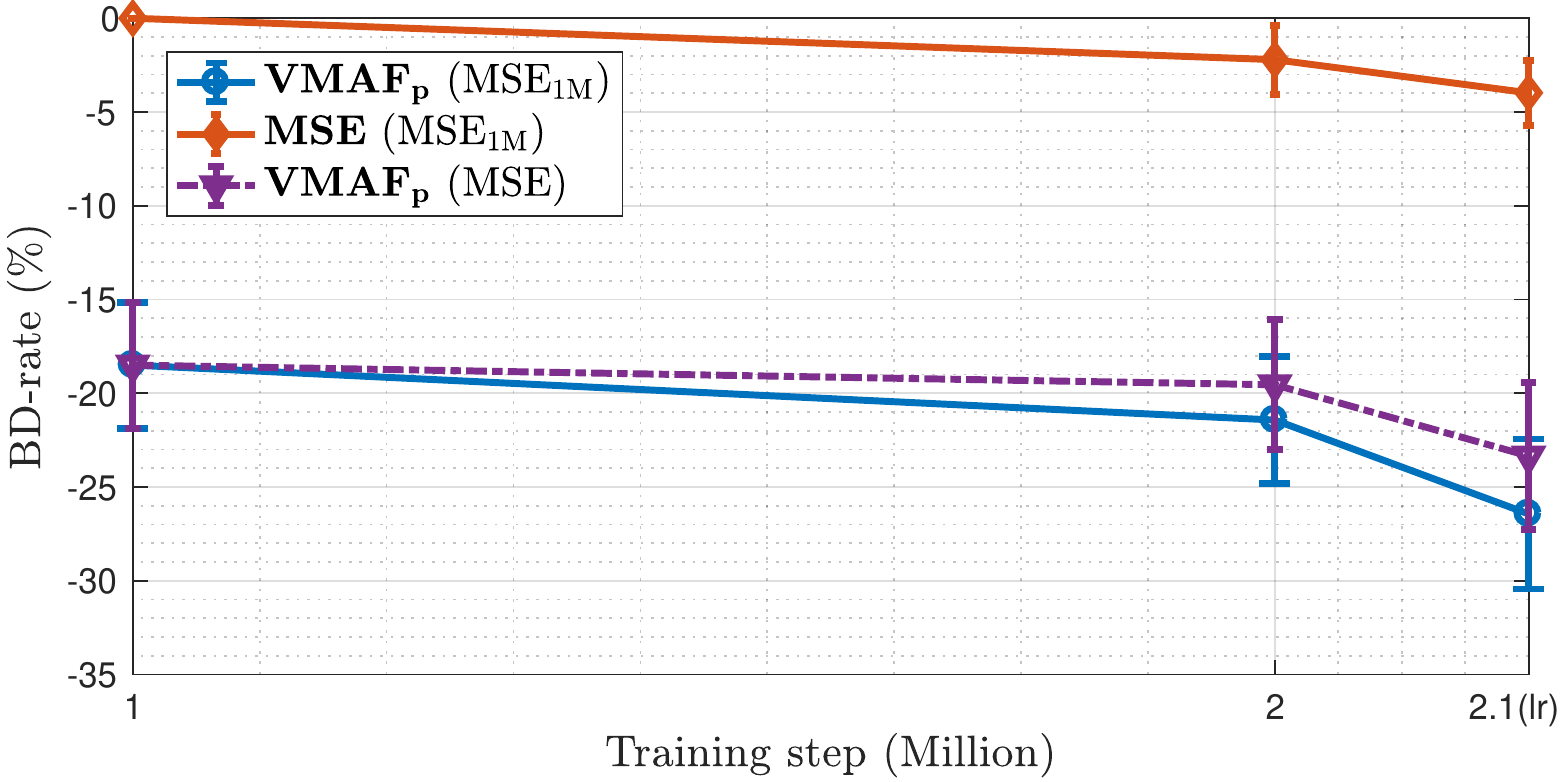}
  \caption{VMAF BD-rate change (improvement) against the number of training steps and learning rate, using the Kodak images. The error bars represent the standard deviations of BD-rates. In the first 2M steps, a constant learning rate was used. After that, the learning rate was reduced by a factor of 10. We denote each result by ``$\mathbf{optimization}$ $\mathrm{(baseline})$". For BD-rate change calculations, the solid lines indicate MSE optimization over 1M steps, while the dotted line indicates that MSE optimization using the same training procedure as the optimization is used for baseline comparison.}
  \label{fig_step_study}
\end{figure}

The instability introduced by the proxy loss can be further improved by training longer, and by reducing the learning rate. Fig.~\ref{fig_step_study} plots the VMAF BD-rate as a function of the number of training steps. When measuring BD-rate against the same baseline (MSE optimization trained with 1M steps), \VMAFp~achieves significant improvement relative to MSE optimization, by training longer or by lowering the learning rate. For fair comparison, we also evaluate \VMAFp~using MSE optimization using the same training process as the baseline (dotted line). We have observed very relative results using other perceptual optimizers, like \SSIMp.

\subsection{Execution Time}
The encoding and decoding times of the various compared codecs are summarized in Table~\ref{tab:time}. We compiled the source code of state-of-the-art (standard) codecs, in order to be able to compare them on the same machine. The results were then calculated by averaging the runtime over all $24$ Kodak images under different bitrate settings. From Table~\ref{tab:time}, it may be observed that the time complexity of the MSE-optimized and \VMAFp-optimized BLS model are nearly identical, as they deploy the same network architecture in application. Furthermore, the encoding processes consume more time than decoding, since the synthesis transform ($g_s$) is also included. Of course, the runtime of deep compression models can be reduced if implemented on a GPU. It should also be noted that the decoding time of HEVC was estimated from the reference software HM, which might be slow. This can be improved by using a third-party decoder such as FFmpeg.

\begin{table}[!t]
  \renewcommand{\arraystretch}{1.3}
  \caption{Run time comparison of conventional image codecs and deep compression models. Model loading time for deep compression is excluded. All times are given in milliseconds.}
  \label{tab:time}
  \centering
  \begin{tabular}{l c r r}
  %\toprule\toprule
  \hline\hline
  Codec & & Encode & Decode \\
  %\midrule
  \hline
  
  JPEG                                & CPU & 43.02   & 62.88          \\
  JPEG2000                            & CPU & 10.80   & 36.79          \\
  HEVC                                & CPU & 4578.57 & 89.88          \\
  \multirow{2}{*}{BLS MSE}            & CPU & 251.01  & 117.93         \\
                                      & \textit{GPU} & \textit{231.62} 
                                      & \textit{32.56}                 \\
  \multirow{2}{*}{BLS \VMAFp ~(ours)} & CPU & 246.57  & 119.02         \\
                                      & \textit{GPU} & \textit{229.26}
                                      & \textit{29.22}                 \\
  %\bottomrule\bottomrule
  \hline\hline
  \end{tabular}
  \end{table}
  
  \begin{figure}[!t]
    \centering
    \includegraphics[width=3.4in]{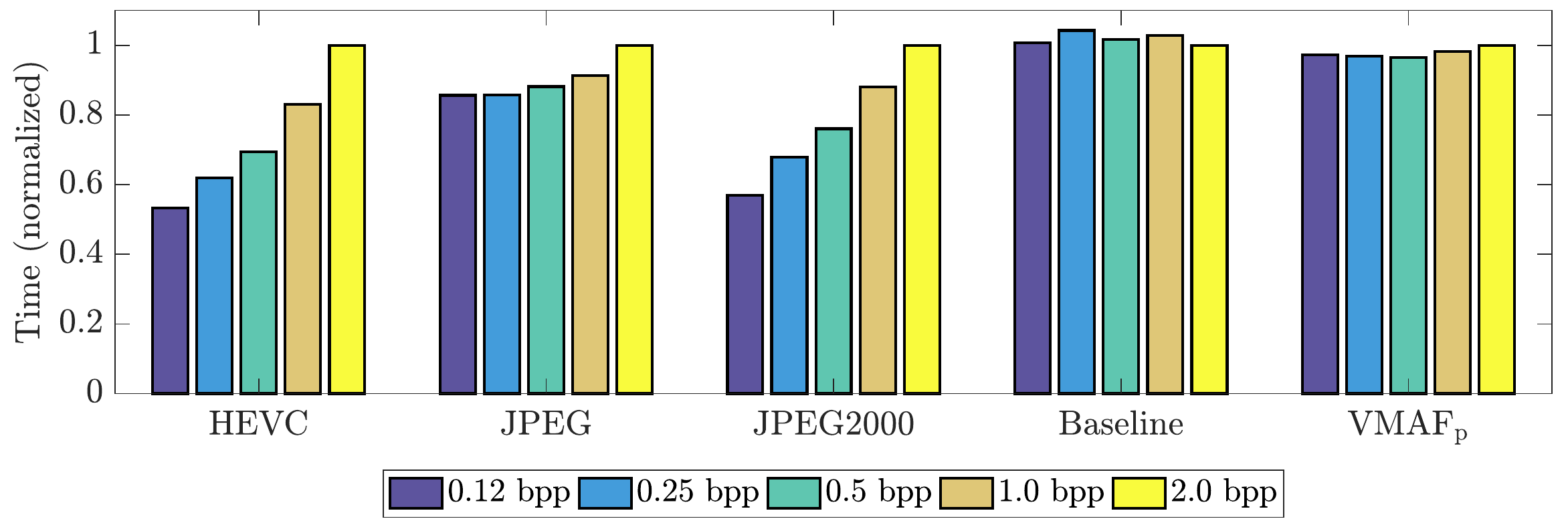}
    \caption{Normalized encoding times for five different bitrate settings.}
    \label{fig_time_bar}
  \end{figure}

We also compared the encoding times for different bitrate settings in Fig.~\ref{fig_time_bar}. For each encoder, all of the runtimes were normalized by dividing by the value of $2.0$ bpp. The conventional codecs required more time to encode at high bitrates, whereas the deep compression models have encoding times that do not vary much with bitrate.

\section{Concluding Remarks}
We have presented a learning framework for perceptually optimizing a learned image compression model. To optimize the ProxIQA network, we developed an alternating training method. We experimentally demonstrated that, for a fixed VMAF value, our proposed proxy approach achieved a $20\%$ bitrate reduction, on average, relative to the MSE-based framework.

The idea behind the proposed optimization framework is general. We believe that, with proper modifications of the architecture of the ProxIQA network, the application scope should be applicable to a wide variety of image enhancement, restoration, and reconstruction problems, such as super-resolution, de-banding \cite{tu2020bband, Tu2020ada}, or de-noising.

Another future topic could be the study of new types of distortions caused by deep compression models. Like the examples shown in this paper, distorted images created by CNNs are very different from images afflicted by more traditional distortions, such as JPEG compression. Creating databases for assessing the subjective quality of these new types of distortions would be quite valuable.

% or


%\appendices
%% you can choose not to have a title for an appendix
%% if you want by leaving the argument blank
%\section{}

\section*{Acknowledgment}
The authors thank Johannes Ball\'e for providing the training images. The authors also thank the Texas Advanced Computing Center (TACC) at The University of Texas at Austin for providing HPC resources that have contributed to the research results reported within this paper. URL: http://www.tacc.utexas.edu

\ifCLASSOPTIONcaptionsoff
  \newpage
\fi



%\bibliographystyle{IEEEtran}

%\begin{thebibliography}{1}
%H.~Kopka and P.~W. Daly, \emph{A Guide to \LaTeX}, 3rd~ed.\hskip 1em %plus
%\end{thebibliography}
\bibliographystyle{IEEEtran}
\bibliography{IEEEabrv,IEEEexample_}

\begin{thebibliography}{10}
\providecommand{\url}[1]{#1}
\csname url@samestyle\endcsname
\providecommand{\newblock}{\relax}
\providecommand{\bibinfo}[2]{#2}
\providecommand{\BIBentrySTDinterwordspacing}{\spaceskip=0pt\relax}
\providecommand{\BIBentryALTinterwordstretchfactor}{4}
\providecommand{\BIBentryALTinterwordspacing}{\spaceskip=\fontdimen2\font plus
\BIBentryALTinterwordstretchfactor\fontdimen3\font minus
  \fontdimen4\font\relax}
\providecommand{\BIBforeignlanguage}[2]{{%
\expandafter\ifx\csname l@#1\endcsname\relax
\typeout{** WARNING: IEEEtran.bst: No hyphenation pattern has been}%
\typeout{** loaded for the language `#1'. Using the pattern for}%
\typeout{** the default language instead.}%
\else
\language=\csname l@#1\endcsname
\fi
#2}}
\providecommand{\BIBdecl}{\relax}
\BIBdecl

\bibitem{ShelhamerLD17}
E.~Shelhamer, J.~Long, and T.~Darrell, ``Fully convolutional networks for
  semantic segmentation,'' \emph{{IEEE} Trans. Pattern Anal. and Mach.
  Intell.}, vol.~39, no.~4, pp. 640--651, Apr. 2017.

\bibitem{HeZRS16}
K.~He, X.~Zhang, S.~Ren, and J.~Sun, ``Deep residual learning for image
  recognition,'' in \emph{Proc. IEEE Conf. Comput. Vision Pattern Recog.}, Jun.
  2016, pp. 770--778.

\bibitem{TMLiu2018}
T.-M. Liu, C.-H. Tsai, T.-H. Wu, J.-Y. Lin, L.-H. Chen, H.-L. Chou, Y.-C.
  Chang, and C.-C. Ju, ``A 0.76 mm2 0.22 {nJ}/pixel {DL}-assisted 4k video
  encoder {LSI} for quality-of-experience over smartphones,'' \emph{{IEEE}
  Solid-State Circuits Lett.}, vol.~1, no.~12, pp. 221--224, 2018.

\bibitem{SPaul2020}
S.~Paul, A.~Norkin, and A.~C. Bovik, ``Speeding up {VP}9 intra encoder with
  hierarchical deep learning-based partition prediction,'' \emph{{IEEE} Trans.
  Image Processing}, vol.~29, pp. 8134--8148, 2020.

\bibitem{Dosovitskiy:2015:FLO}
A.~Dosovitskiy, P.~Fischer, E.~Ilg, P.~Hausser, C.~Hazirbas, V.~Golkov,
  P.~van~der Smagt, D.~Cremers, and T.~Brox, ``{FlowNet}: Learning optical flow
  with convolutional networks,'' in \emph{Proc. {IEEE} Int. Conf. Comput.
  Vision}, Dec. 2015, pp. 2758--2766.

\bibitem{BurgerSH12}
H.~C. Burger, C.~J. Schuler, and S.~Harmeling, ``Image denoising: Can plain
  neural networks compete with {BM3D}?'' in \emph{Proc. IEEE Conf. Comput.
  Vision Pattern Recog.}, Jun. 2012, pp. 2392--2399.

\bibitem{NIPS2008_3506}
V.~Jain and S.~Seung, ``Natural image denoising with convolutional networks,''
  in \emph{Proc. Adv. Neural Inf. Process. Syst.}, 2009, pp. 769--776.

\bibitem{dong2014learning}
C.~Dong, C.~C. Loy, K.~He, and X.~Tang, ``Learning a deep convolutional network
  for image super-resolution,'' in \emph{Proc. Eur. Conf. Comput. Vision},
  2014, pp. 184--199.

\bibitem{MSLapSRN}
W.-S. Lai, J.-B. Huang, N.~Ahuja, and M.-H. Yang, ``Deep laplacian pyramid
  networks for fast and accurate super-resolution,'' in \emph{Proc. IEEE Conf.
  Comput. Vision Pattern Recog.}, Jun. 2017, pp. 624--632.

\bibitem{liu2019cyclicgen}
Y.-L. Liu, Y.-T. Liao, Y.-Y. Lin, and Y.-Y. Chuang, ``Deep video frame
  interpolation using cyclic frame generation,'' in \emph{Proc. AAAI}, 2019,
  pp. 8794--8802.

\bibitem{BarronCVPR2019}
J.~T. Barron, ``A general and adaptive robust loss function,'' in \emph{Proc.
  IEEE Conf. Comput. Vision Pattern Recog.}, Jun. 2019, pp. 4331--4339.

\bibitem{WangBSS04}
Z.~Wang, A.~Bovik, H.~Sheikh, and E.~Simoncelli, ``Image quality assessment:
  From error visibility to structural similarity,'' \emph{IEEE Trans. Image
  Processing}, vol.~13, no.~4, pp. 600--612, Apr. 2004.

\bibitem{WangMSSSIM03}
Z.~Wang, E.~P. Simoncelli, and A.~C. Bovik, ``Multi-scale structural similarity
  for image quality assessment,'' in \emph{Proc. IEEE Asilomar Conf. on
  Signals, Syst., and Comput.}, Nov. 2003, pp. 1398--1402.

\bibitem{Snell2017}
J.~Snell, K.~Ridgeway, R.~Liao, B.~D. Roads, M.~C. Mozer, and R.~S. Zemel,
  ``Learning to generate images with perceptual similarity metrics,'' in
  \emph{Proc. {IEEE} Int. Conf. Image Process.}, Sep. 2017, pp. 4277--4281.

\bibitem{Zhao2017}
H.~Zhao, O.~Gallo, I.~Frosio, and J.~Kautz, ``Loss functions for image
  restoration with neural networks,'' \emph{{IEEE} Trans. on Computational
  Imaging}, vol.~3, no.~1, pp. 47--57, Mar. 2017.

\bibitem{Wan2020}
Z.~Wan, K.~Gu, and D.~Zhao, ``Reduced reference stereoscopic image quality
  assessment using sparse representation and natural scene statistics,''
  \emph{{IEEE} Trans. Multimedia}, vol.~22, no.~8, pp. 2024--2037, Aug. 2020.

\bibitem{ztuugcvqa2020}
Z.~Tu, Y.~Wang, N.~Birkbeck, B.~Adsumilli, and A.~C. Bovik, ``{UGC-VQA}:
  Benchmarking blind video quality assessment for user generated content,''
  \emph{arXiv preprint arXiv:2005.14354}, 2020.

\bibitem{DMCSSHVSNR07}
D.~Chandler and S.~Hemami, ``{VSNR}: A wavelet-based visual signal-to-noise
  ratio for natural images,'' \emph{IEEE Trans. Image Processing}, vol.~16,
  no.~9, pp. 2284--2298, Sep. 2007.

\bibitem{SheikhB06}
H.~R. Sheikh and A.~C. Bovik, ``Image information and visual quality,''
  \emph{IEEE Trans. Image Processing}, vol.~15, no.~2, pp. 430--444, Feb. 2006.

\bibitem{Chandler2010}
E.~C. Larson and D.~M. Chandler, ``Most apparent distortion: full-reference
  image quality assessment and the role of strategy,'' \emph{Journal of
  Electronic Imaging}, vol.~19, no.~1, pp. 011\,006:1--011\,006:21, Jan. 2010.

\bibitem{LZhangFSIM2011}
L.~Zhang, L.~Zhang, X.~Mou, and D.~Zhang, ``{FSIM}: A feature similarity index
  for image quality assessment,'' \emph{IEEE Trans. Image Processing}, vol.~20,
  no.~8, pp. 2378--2386, Aug. 2011.

\bibitem{ZhangSLVSI14}
L.~Zhang, Y.~Shen, and H.~Li, ``{VSI}: A visual saliency-induced index for
  perceptual image quality assessment,'' \emph{IEEE Trans. Image Processing},
  vol.~23, no.~10, pp. 4270--4281, Oct. 2014.

\bibitem{Liu2013}
T.-J. Liu, W.~Lin, and C.-C.~J. Kuo, ``Image quality assessment using
  multi-method fusion,'' \emph{IEEE Trans. Image Processing}, vol.~22, no.~5,
  pp. 1793--1807, May 2013.

\bibitem{PeiDOG15}
S.-C. Pei and L.-H. Chen, ``Image quality assessment using human visual {DOG}
  model fused with random forest,'' \emph{IEEE Trans. Image Processing},
  vol.~24, no.~11, pp. 3282--3292, Nov. 2015.

\bibitem{LukinPIEA15}
V.~V. Lukin, N.~N. Ponomarenko, O.~I. Ieremeiev, K.~O. Egiazarian, and
  J.~Astola, ``Combining full-reference image visual quality metrics by neural
  network,'' \emph{Proc. SPIE}, vol. 9394, p. 93940K, Mar. 2015.

\bibitem{Oszust2016}
M.~Oszust, ``Decision fusion for image quality assessment using an optimization
  approach,'' \emph{{IEEE} Signal Process. Lett.}, vol.~23, no.~1, pp. 65--69,
  Jan. 2016.

\bibitem{GaoWLTYZ17}
F.~Gao, Y.~Wang, P.~Li, M.~Tan, J.~Yu, and Y.~Zhu, ``{DeepSim}: Deep similarity
  for image quality assessment,'' \emph{Neurocomputing}, vol. 257, pp.
  104--114, Sep. 2017.

\bibitem{Bosse2018}
S.~Bosse, D.~Maniry, K.-R. Muller, T.~Wiegand, and W.~Samek, ``Deep neural
  networks for no-reference and full-reference image quality assessment,''
  \emph{IEEE Trans. Image Processing}, vol.~27, no.~1, pp. 206--219, Jan. 2018.

\bibitem{Bosse2019dsp}
S.~Bosse, S.~Becker, K.-R. M\"{u}ller, W.~Samek, and T.~Wiegand, ``Estimation
  of distortion sensitivity for visual quality prediction using a convolutional
  neural network,'' \emph{Digit. Signal Process.}, vol.~91, pp. 54--65, Aug.
  2019.

\bibitem{ZliVMAF18}
\BIBentryALTinterwordspacing
Z.~Li, C.~Bampis, J.~Novak, A.~Aaron, K.~Swanson, A.~Moorthy, and J.~D. Cock,
  ``{VMAF}: The journey continues,'' \emph{The NETFLIX tech blog}, 2018.
  [Online]. Available:
  \url{https://netflixtechblog.com/vmaf-the-journey-continues-44b51ee9ed12}
\BIBentrySTDinterwordspacing

\bibitem{Sinno2020}
Z.~Sinno, A.~K. Moorthy, J.~D. Cock, Z.~Li, and A.~C. Bovik, ``Quality
  measurement of images on mobile streaming interfaces deployed at scale,''
  \emph{{IEEE} Trans. Image Process.}, vol.~29, pp. 2536--2551, 2019.

\bibitem{Brunet2012}
D.~Brunet, E.~R. Vrscay, and Z.~Wang, ``On the mathematical properties of the
  structural similarity index,'' \emph{IEEE Trans. Image Processing}, vol.~21,
  no.~4, pp. 1488--1499, Apr. 2012.

\bibitem{dingperceptualiqa2020}
K.~Ding, K.~Ma, S.~Wang, and E.~P. Simoncelli, ``Comparison of image quality
  models for optimization of image processing systems,'' \emph{arXiv preprint
  arXiv:2005.01338}, 2020.

\bibitem{zhang2018perceptual}
R.~Zhang, P.~Isola, A.~A. Efros, E.~Shechtman, and O.~Wang, ``The unreasonable
  effectiveness of deep features as a perceptual metric,'' in \emph{Proc. IEEE
  Conf. Comput. Vision Pattern Recog.}, Jun. 2018, pp. 586--595.

\bibitem{SimonyanZ14a}
K.~Simonyan and A.~Zisserman, ``Very deep convolutional networks for
  large-scale image recognition,'' in \emph{Proc. Int. Conf. Learn.
  Represent.}, 2015, pp. 1--14.

\bibitem{JohnsonAF16}
J.~Johnson, A.~Alahi, and L.~Fei{-}Fei, ``Perceptual losses for real-time style
  transfer and super-resolution,'' in \emph{Proc. Eur. Conf. Comput. Vision},
  vol. 9906, 2016, pp. 694--711.

\bibitem{GatysEBHS17}
L.~A. Gatys, A.~S. Ecker, M.~Bethge, A.~Hertzmann, and E.~Shechtman,
  ``Controlling perceptual factors in neural style transfer,'' in \emph{Proc.
  IEEE Conf. Comput. Vision Pattern Recog.}, Jul. 2017, pp. 3730--3738.

\bibitem{BrunaSL15}
J.~Bruna, P.~Sprechmann, and Y.~LeCun, ``Super-resolution with deep
  convolutional sufficient statistics,'' in \emph{Proc. Int. Conf. Learn.
  Represent.}, 2016, pp. 1--17.

\bibitem{LedigTHCCAATTWS17}
C.~Ledig, L.~Theis, F.~Huszar, J.~Caballero, A.~Cunningham, A.~Acosta, A.~P.
  Aitken, A.~Tejani, J.~Totz, Z.~Wang, and W.~Shi, ``Photo-realistic single
  image super-resolution using a generative adversarial network,'' in
  \emph{Proc. IEEE Conf. Comput. Vision Pattern Recog.}, 2017, pp. 105--114.

\bibitem{Sajjadi2017}
M.~S.~M. Sajjadi, B.~Scholkopf, and M.~Hirsch, ``{EnhanceNet}: Single image
  super-resolution through automated texture synthesis,'' in \emph{{Proc.
  {IEEE} Int. Conf. Comput. Vision}}, Oct. 2017.

\bibitem{Yang_2017_CVPR}
C.~Yang, X.~Lu, Z.~Lin, E.~Shechtman, O.~Wang, and H.~Li, ``High-resolution
  image inpainting using multi-scale neural patch synthesis,'' in \emph{Proc.
  IEEE Conf. Comput. Vision Pattern Recog.}\hskip 1em plus 0.5em minus
  0.4em\relax {IEEE}, Jul. 2017, pp. 4076--4084.

\bibitem{BalleLS16a}
J.~Ball\'{e}, V.~Laparra, and E.~P. Simoncelli, ``End-to-end optimized image
  compression,'' in \emph{Proc. Int. Conf. Learn. Represent.}, 2017, pp. 1--27.

\bibitem{balle2018variational}
J.~Ball\'{e}, D.~Minnen, S.~Singh, S.~J. Hwang, and N.~Johnston, ``Variational
  image compression with a scale hyperprior,'' in \emph{Proc. Int. Conf. Learn.
  Represent.}, 2018, pp. 1--23.

\bibitem{NIPS2018_8275}
D.~Minnen, J.~Ball\'{e}, and G.~D. Toderici, ``Joint autoregressive and
  hierarchical priors for learned image compression,'' in \emph{Advances in
  Neural Information Processing Systems 31}, 2018, pp. 10\,771--10\,780.

\bibitem{Toderici2015VariableRI}
G.~Toderici, S.~M. O'Malley, S.~J. Hwang, D.~Vincent, D.~Minnen, S.~Baluja,
  M.~Covell, and R.~Sukthankar, ``Variable rate image compression with
  recurrent neural networks,'' \emph{CoRR}, vol. abs/1511.06085, 2015.

\bibitem{Toderici2017}
G.~Toderici, D.~Vincent, N.~Johnston, S.~J. Hwang, D.~Minnen, J.~Shor, and
  M.~Covell, ``Full resolution image compression with recurrent neural
  networks,'' in \emph{Proc. IEEE Conf. Comput. Vision Pattern Recog.}, Jul.
  2017, pp. 5306--5314.

\bibitem{Johnston_2018_CVPR}
N.~Johnston, D.~Vincent, D.~Minnen, M.~Covell, S.~Singh, T.~Chinen,
  S.~Jin~Hwang, J.~Shor, and G.~Toderici, ``Improved lossy image compression
  with priming and spatially adaptive bit rates for recurrent networks,'' in
  \emph{Proc. IEEE Conf. Comput. Vision Pattern Recog.}, June 2018, pp.
  4385--4393.

\bibitem{agustsson2018generative}
E.~Agustsson, M.~Tschannen, F.~Mentzer, R.~Timofte, and L.~Van~Gool,
  ``Generative adversarial networks for extreme learned image compression,'' in
  \emph{Proc. {IEEE} Int. Conf. Comput. Vision}, Oct. 2019, pp. 221--231.

\bibitem{Lhdefink2019GANVJ}
J.~L{\"o}hdefink, A.~B{\"a}r, N.~M. Schmidt, F.~H{\"u}ger, P.~Schlicht, and
  T.~Fingscheidt, ``{GAN}- vs. {JPEG}2000 image compression for distributed
  automotive perception: Higher peak snr does not mean better semantic
  segmentation,'' \emph{arXiv preprint arXiv:1902.04311}, 2019.

\bibitem{wu2018vcii}
C.-Y. Wu, N.~Singhal, and P.~Kr{\"a}henb{\"u}hl, ``Video compression through
  image interpolation,'' in \emph{Proc. Eur. Conf. Comput. Vision}, 2018.

\bibitem{cheng19}
Z.~Cheng, H.~Sun, M.~Takeuchi, and J.~Katto, ``Learning image and video
  compression through spatial-temporal energy compaction,'' in \emph{Proc. IEEE
  Conf. Comput. Vision Pattern Recog.}, 2019.

\bibitem{kimvideo}
S.~Kim, J.~S. Park, C.~G. Bampis, J.~Lee, M.~K. Markey, A.~G. Dimakis, and
  A.~C. Bovik, ``Adversarial video compression guided by soft edge detection,''
  in \emph{Proc. IEEE Int. Conf. Acoust., Speech and Signal Process. (ICASSP)},
  2020, pp. 2193--2197.

\bibitem{channappayya08}
S.~Channappayya, A.~Bovik, and R.~Heath, ``Rate bounds on {SSIM} index of
  quantized images,'' \emph{{IEEE} Trans. Image Processing}, vol.~17, no.~9,
  pp. 1624--1639, Sep. 2008.

\bibitem{YHH10}
Y.-H. Huang, T.-S. Ou, P.-Y. Su, and H.~H. Chen, ``Perceptual rate-distortion
  optimization using structural similarity index as quality metric,''
  \emph{{IEEE} Trans. Circuits Syst. Video Technol.}, vol.~20, no.~11, pp.
  1614--1624, Nov. 2010.

\bibitem{WangRWMG12}
S.~Wang, A.~Rehman, Z.~Wang, S.~Ma, and W.~Gao, ``{SSIM}-motivated
  rate-distortion optimization for video coding,'' \emph{{IEEE} Trans. Circuits
  Syst. Video Technol.}, vol.~22, no.~4, pp. 516--529, 2012.

\bibitem{kslussim20}
K.-S. Lu, A.~Ortega, D.~Mukherjee, and Y.~Chen, ``Perceptually inspired
  weighted mse optimization using irregularity-aware graph fourier transform,''
  in \emph{Proc. {IEEE} Int. Conf. Image Process.}, Sep. 2020, pp. 3384--3388.

\bibitem{NIPSGAN}
I.~Goodfellow, J.~Pouget-Abadie, M.~Mirza, B.~Xu, D.~Warde-Farley, S.~Ozair,
  A.~Courville, and Y.~Bengio, ``Generative adversarial nets,'' in \emph{Proc.
  Adv. Neural Inf. Process. Syst.}, 2014, pp. 2672--2680.

\bibitem{pcs_Balle18}
J.~Ball{\'{e}}, ``Efficient nonlinear transforms for lossy image compression,''
  in \emph{Proc. IEEE Picture Coding Symp.}, 2018, pp. 248--252.

\bibitem{cabac_tcsvt03}
D.~Marpe, H.~Schwarz, and T.~Wiegand, ``Context-based adaptive binary
  arithmetic coding in the h.264/{AVC} video compression standard,''
  \emph{{IEEE} Trans. Circuits Syst. Video Technol.}, vol.~13, no.~7, pp.
  620--636, Jul. 2003.

\bibitem{ma2017waterloo}
K.~Ma, Z.~Duanmu, Q.~Wu, Z.~Wang, H.~Yong, H.~Li, and L.~Zhang, ``{Waterloo
  Exploration Database}: New challenges for image quality assessment models,''
  \emph{IEEE Trans. Image Processing}, vol.~26, no.~2, pp. 1004--1016, Feb.
  2017.

\bibitem{kingma:adam}
D.~P. Kingma and J.~Ba, ``Adam: A method for stochastic optimization,'' in
  \emph{Proc. Int. Conf. Learn. Represent.}, 2015, pp. 1--15.

\bibitem{imagenet_cvpr09}
J.~Deng, W.~Dong, R.~Socher, L.-J. Li, K.~Li, and L.~Fei-Fei, ``{ImageNet}: A
  large-scale hierarchical image database,'' in \emph{Proc. IEEE Conf. Comput.
  Vision Pattern Recog.}, Jun. 2009, pp. 248--255.

\bibitem{kodak_data}
\BIBentryALTinterwordspacing
\emph{Kodak lossless true color image suite}, 2007. [Online]. Available:
  \url{http://r0k.us/graphics/kodak/}
\BIBentrySTDinterwordspacing

\bibitem{stag_tecnick}
N.~Asuni and A.~Giachetti, ``{TESTIMAGES}: a large-scale archive for testing
  visual devices and basic image processing algorithms,'' in \emph{Proc.
  Eurographics Italian Chapter Conference}, 2014, pp. 63--70.

\bibitem{BDRate01}
{G. Bj\o{}ntegaard}, ``Calculation of average {PSNR} differences between
  {RD}-curves,'' \emph{document VCEG-M33, ITU-T Video Coding Experts Group
  (VCEG) Thirteenth Meeting}, Austin, TX, April 2001.

\bibitem{Blau2018}
Y.~Blau and T.~Michaeli, ``The perception-distortion tradeoff,'' in \emph{Proc.
  IEEE Conf. Comput. Vision Pattern Recog.}, Jun. 2018.

\bibitem{btscore1952}
R.~A. Bradley and M.~E. Terry, ``Rank analysis of incomplete block designs: The
  method of paired comparisons,'' \emph{Biometrika}, vol.~39, no. 3-4, pp.
  324--345, Dec. 1952.

\bibitem{tu2020bband}
Z.~{Tu}, J.~{Lin}, Y.~{Wang}, B.~{Adsumilli}, and A.~C. {Bovik}, ``{BBAND}
  index: A no-reference banding artifact predictor,'' in \emph{Proc. IEEE Int.
  Conf. Acoust., Speech and Signal Process. (ICASSP)}, 2020, pp. 2712--2716.

\bibitem{Tu2020ada}
Z.~Tu, J.~Lin, Y.~Wang, B.~Adsumilli, and A.~C. Bovik, ``Adaptive debanding
  filter,'' \emph{{IEEE} Signal Process. Lett.}, vol.~27, pp. 1715--1719, 2020.

\end{thebibliography}

\end{document}